\documentclass[10pt,twocolumn]{article}

\usepackage[a4paper,margin=0.85in,columnsep=0.28in]{geometry}
\usepackage{balance}
\usepackage{amsmath,amssymb,amsthm}
\usepackage{graphicx}
\usepackage{booktabs}
\usepackage{tikz}
\usepackage{pgfplots}
\pgfplotsset{compat=1.17}
\usetikzlibrary{positioning,fit,backgrounds,calc,arrows.meta}
\usetikzlibrary{quantikz2}
\usepackage{xcolor}
\usepackage[hidelinks]{hyperref}
\usepackage{caption}
\usepackage{algorithm}
\usepackage{algpseudocode}

\usepackage[numbers,sort&compress]{natbib}
\usepackage{etoolbox}
\patchcmd{\thebibliography}{\section*{\refname}}{\section*{\refname}\footnotesize}{}{}

\renewcommand{\ket}[1]{\lvert #1 \rangle}
\renewcommand{\bra}[1]{\langle #1 \rvert}
\newcommand{\ketbra}[2]{\lvert #1 \rangle\!\langle #2 \rvert}
\renewcommand{\braket}[2]{\langle #1 \vert #2 \rangle}
\newcommand{\dq}{\mathrm{D}}
\newcommand{\aq}{\mathrm{A}}
\newcommand{\RY}{R_Y}
\newcommand{\RZ}{R_Z}
\definecolor{dataclr}{RGB}{31,119,180}
\definecolor{addrclr}{RGB}{214,110,20}

\newif\ifrevision \revisiontrue

\newtheorem{proposition}{Proposition}

\usepackage{flushend}
\usepackage{microtype}
\title{Tessa: Encoding Classical Data Across an Entire Quantum Processor}
\author{Daniel Bultrini\\ \small Moth Quantum \\ \small\texttt{daniel@mothquantum.com}}
\date{\today}

\begin{document}

\makeatletter
\twocolumn[\begin{@twocolumnfalse}
\maketitle
\begin{quote}
\begin{center}\textbf{Abstract}\end{center}\vspace{-2pt}
\small

We present Tessa, a hardware-aware, device-wide scheme to encode classical data on quantum computers. We use it to experimentally encode and decode up to $1568$ real values using IBM's Phoenix and Fez quantum computers on a variety of data. Tessa's construction is an extension of QPIXL and QCrank. However, it checkerboards address and data qubits across a coupling graph with neighbouring data qubits sharing only a subset of address qubits. When using the device's coupling graph no SWAP gates or routing is required, but the construction works for any bipartite coupling graph. The shots required for readout are determined by the local coupling degree.  As examples, we have used cat states (pixel art of cats), various pixel art, audio, MIDI, DNA sequences and protein amino acid sequences.

\end{quote}
\vspace{1.2em}
\end{@twocolumnfalse}]
\makeatother
\section{Introduction}

\begin{figure}[!b]
\centering\small
\setlength{\tabcolsep}{1pt}
\begin{tabular}{ccc}
& Input & Hardware \\
\rotatebox[origin=c]{90}{\parbox{0.33\linewidth}{\centering\footnotesize QCAT\\$784$ values}} &
\raisebox{-0.5\height}{\includegraphics[width=0.33\linewidth,trim=22 129.6 399.3 14.7,clip]{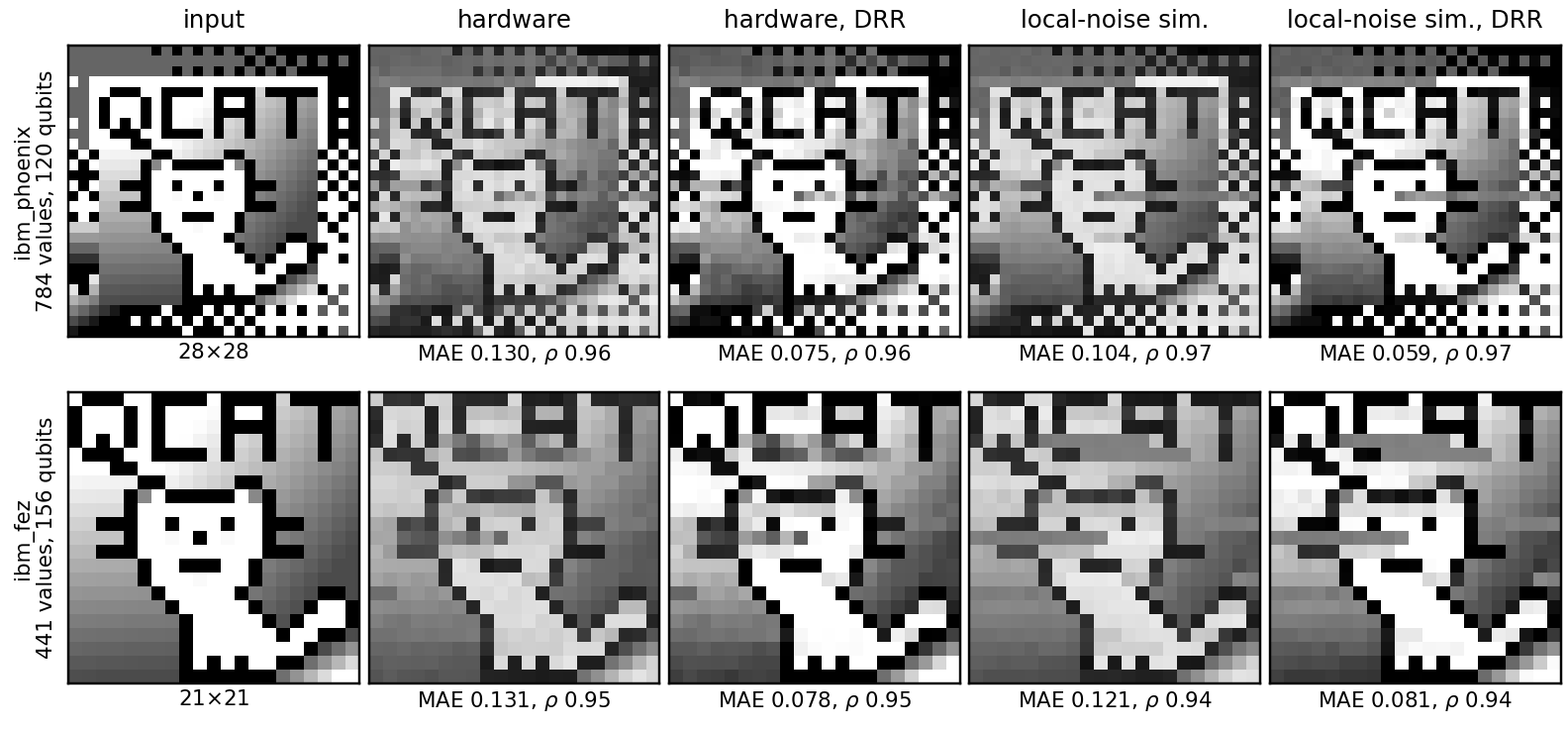}} &
\raisebox{-0.5\height}{\includegraphics[width=0.33\linewidth,trim=121.7 129.6 300.1 14.7,clip]{figures/fig_qcat_hardware_z.png}} \\[2pt]
\rotatebox[origin=c]{90}{\parbox{0.33\linewidth}{\centering\footnotesize City-pop cat\\$1568$ values}} &
\raisebox{-0.5\height}{\includegraphics[width=0.33\linewidth,trim=276.7 347.7 132.5 37,clip]{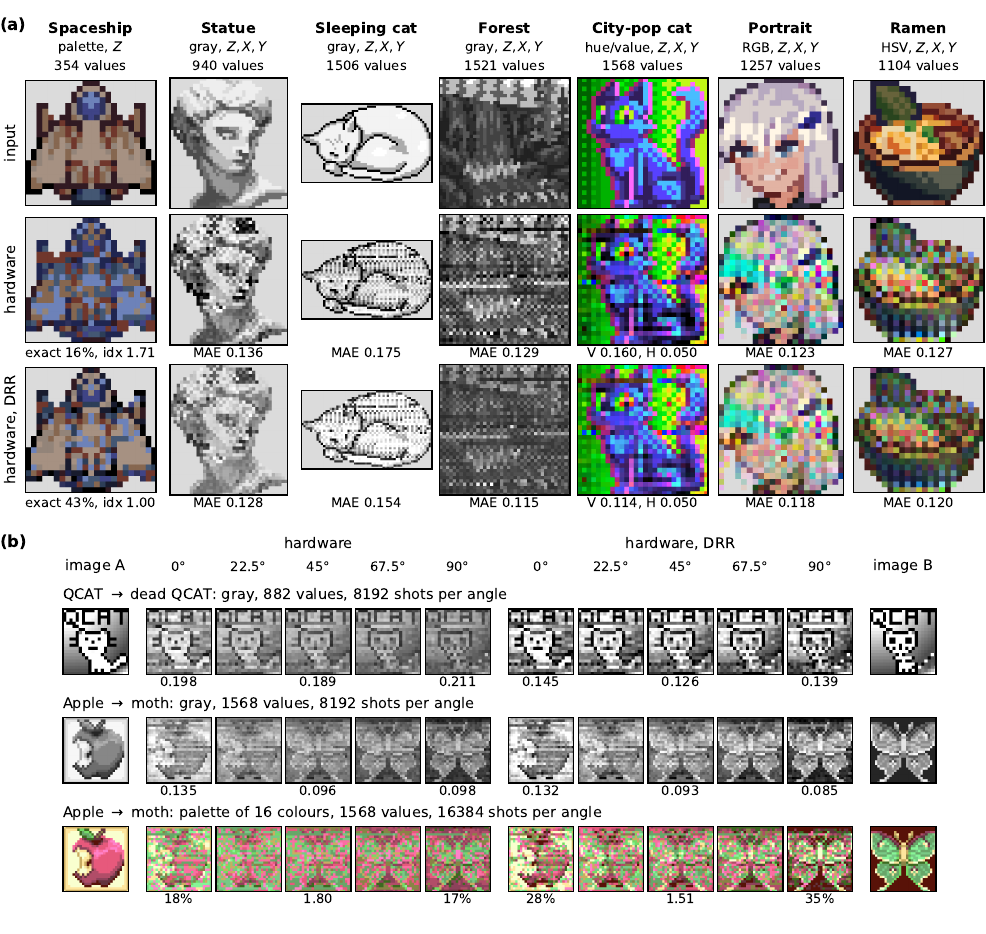}} &
\raisebox{-0.5\height}{\includegraphics[width=0.33\linewidth,trim=276.7 282.1 132.5 102.6,clip]{figures/fig_pixelart_hardware_raw.pdf}}\\
\end{tabular}
\caption{Pixel art encoded with Tessa on IBM's Phoenix quantum computer and decoded from $8192$ shots per basis. Top: QCAT, one value per pixel encoded in $Z$. Bottom: City-pop cat by M. Fish (Leviatic), value and hue, two values per pixel encoded in $Z$ and $(X,Y)$.}
\label{fig:teaser}
\end{figure}

Encoding classical data onto a quantum computer is one of the known bottlenecks for quantum machine learning on classical data, which is one of the areas where quantum advantage is believed to lie~\cite{zhao2026exponentialquantumadvantageprocessing}. But performing this encoding is a fundamental problem known as the \textit{data encoding problem}~\cite{lloyd2020quantumembeddingsmachinelearning}. There is a rich tapestry of methods that have been developed to address this, as it is the first step in quantum machine learning to process classical data on a quantum computer.
We focus on the amplitude and angle encoding approach, such as FRQI~\cite{frqi} and QPIXL~\cite{qpixl}. These write a vector of length $N$ into $\log_2 N$ address qubits and one data qubit with a uniformly controlled rotation, called a multiplexer: a rotation of the data qubit whose angle depends on the computational basis state of the address qubits. QCrank~\cite{qcrank} parallelised the encoding by adding more data qubits that share one group of address qubits.
The aforementioned schemes require a star connectivity from the data to address qubits, which requires expensive operations on hardware with limited connectivity. 

Two other lines of earlier work are related: divide-and-conquer state preparation~\cite{araujo2021} and low-depth state preparation~\cite{zhang2022}. These methods reduce depth by adding ancilla qubits, or by splitting a preparation across disjoint registers. Tessa needs no ancilla qubits, because its parallelism comes from the geometry of the device. NEQR~\cite{zhang2013neqr} and related encodings store each value in a register in basis encoding. These encodings are also exact, but they use many more qubits. We focus on angle encoding, in which each value sets a rotation angle, and the number of shots sets the precision of the recovered value.

All-to-all hardware connectivity does not solve this problem fully. Trapped-ion~\cite{pino2021} and neutral-atom~\cite{bluvstein2022} processors offer non-local gates, but these gates are usually slower, or they require moving the qubits. An encoding that works with a connectivity graph of our choice, with fewer long-range gates, can therefore also help on these machines.

\begin{figure}[t]
\centering
\colorlet{gentlered}{red!40}
\begin{tikzpicture}[scale=0.82,
  dnode/.style={circle,draw,thick,fill=dataclr!25,minimum size=6mm,inner sep=0pt,font=\small},
  anode/.style={circle,draw,thick,fill=addrclr!25,minimum size=6mm,inner sep=0pt,font=\small},
  eR/.style={black,very thick},
  eL/.style={green!45!black,very thick,dashed},
  eD/.style={cyan!60!black,very thick,dotted},
  eU/.style={orange!70!black,very thick,dash dot}]
  \foreach \r in {0,...,3} {
    \foreach \c in {0,...,3} {
      \coordinate (n-\r-\c) at (\c*1.6, -\r*1.6);
    }
  }
  \foreach \r in {0,...,3} {
    \foreach \c in {0,1,2} {
      \pgfmathtruncatemacro{\cc}{\c+1}
      \pgfmathtruncatemacro{\isdata}{mod(\r+\c,2)==0 ? 1 : 0}
      \ifnum\isdata=1 \draw[eR] (n-\r-\c)--(n-\r-\cc);
      \else \draw[eL] (n-\r-\c)--(n-\r-\cc); \fi
    }
  }
  \foreach \r in {0,1,2} {
    \foreach \c in {0,...,3} {
      \pgfmathtruncatemacro{\rr}{\r+1}
      \pgfmathtruncatemacro{\isdata}{mod(\r+\c,2)==0 ? 1 : 0}
      \ifnum\isdata=1 \draw[eD] (n-\r-\c)--(n-\rr-\c);
      \else \draw[eU] (n-\r-\c)--(n-\rr-\c); \fi
    }
  }
  \begin{scope}[on background layer]
    \foreach \r in {0,...,3} {
      \foreach \c in {0,...,3} {
        \pgfmathtruncatemacro{\isdata}{mod(\r+\c,2)==0 ? 1 : 0}
        \ifnum\isdata=1
          \pgfmathtruncatemacro{\isblue}{mod(\r,2)==1 ? 1 : 0}
          \ifnum\isblue=1
            \colorlet{tileclr}{dataclr}
          \else
            \colorlet{tileclr}{gentlered}
          \fi
          \ifnum\r>0 \pgfmathtruncatemacro{\up}{\r-1} \coordinate (top) at (n-\up-\c);
          \else \coordinate (top) at (n-\r-\c); \fi
          \ifnum\r<3 \pgfmathtruncatemacro{\dn}{\r+1} \coordinate (bot) at (n-\dn-\c);
          \else \coordinate (bot) at (n-\r-\c); \fi
          \ifnum\c>0 \pgfmathtruncatemacro{\lf}{\c-1} \coordinate (lft) at (n-\r-\lf);
          \else \coordinate (lft) at (n-\r-\c); \fi
          \ifnum\c<3 \pgfmathtruncatemacro{\rt}{\c+1} \coordinate (rgt) at (n-\r-\rt);
          \else \coordinate (rgt) at (n-\r-\c); \fi
          \filldraw[tileclr, fill opacity=0.25, draw opacity=0.6, thick, line join=round]
            (top) -- (rgt) -- (bot) -- (lft) -- cycle;
        \fi
      }
    }
  \end{scope}
  \foreach \r in {0,...,3} {
    \foreach \c in {0,...,3} {
      \pgfmathtruncatemacro{\isdata}{mod(\r+\c,2)==0 ? 1 : 0}
      \ifnum\isdata=1
        \node[dnode] at (n-\r-\c) {$\dq$};
      \else
        \node[anode] at (n-\r-\c) {$\aq$};
      \fi
    }
  }
\end{tikzpicture}
\caption{The checkerboard lattice ($4\times4$ for illustrative purposes). Data qubits ($\dq$) carry the encoded angles; address qubits ($\aq$) form the shared control register. Each tinted plaquette joins a data qubit to its address neighbours.
An interior data qubit, such as the one in row $1$ and column $1$, has four address neighbours, and together with them it encodes $2^4=16$ values. The line style of each edge gives its direction as seen from its data qubit (right: solid, left: dashed, down: dotted, up: dash-dotted).}
\label{fig:lattice}
\end{figure}
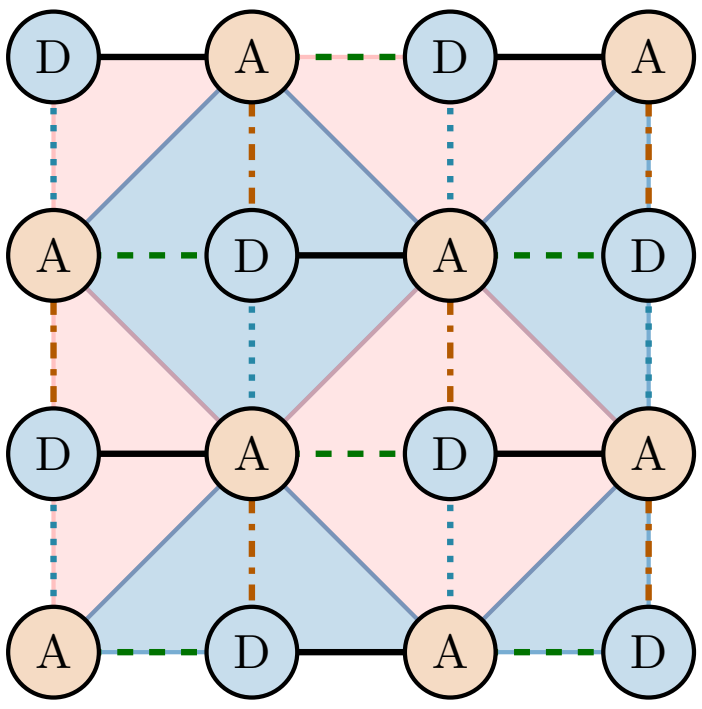

Our solution to this is Tessellated QPIXL (Tessa for short), so called because it fills the device with overlapping QPIXL blocks that share address qubits like in QCrank~\cite{qcrank}, but differently. The construction begins by 2-colouring a desired coupling graph: we label each qubit as an address ($\aq$) or a data ($\dq$) qubit, such that every coupler joins a data qubit to an address qubit (Fig.~\ref{fig:lattice}). Each data qubit is the target of a multiplexer controlled by its neighbouring address qubits which are themselves shared. The construction works on any bipartite coupling graph, but using a device's physical layout results in a circuit without SWAP gates. This allowed us to use all the qubits of IBM's Phoenix device to encode $28\times28$ images in grayscale and in colour, using only hue and value as shown in Fig.\ref{fig:teaser}.

The paper is structured as follows. Section~\ref{sec:lattice} gives the background on QPIXL and QCrank and defines Tessa. Section~\ref{sec:hardware} presents and discusses the hardware results. Section~\ref{sec:conclusion} is the conclusion. The appendices give the optimal circuit construction (App.~\ref{sec:depth}) full proofs (App.~\ref{app:proof}), the entanglement of the encoded state (App.~\ref{app:entanglement}), a computation on the encoded state (App.~\ref{sec:downstream}), full single qubit tomography readout (App.~\ref{sec:xyz}), classical corrections (App.~\ref{sec:considerations}), and further experiments (App.~\ref{app:music}), including music, audio, DNA sequences, and protein amino acid sequences such as p53~\citep{Lane1992}, Pikachurin~\citep{Sato2008} and Sonic Hedgehog~\citep{Riddle1993}.

\section{Tessellated QPIXL}
\label{sec:lattice}

\subsection{QPIXL and QCrank}
\label{sec:background}

 QPIXL stores a vector of $2^k$ values with $k$ address qubits and one data qubit. Let $a=(a_0,\dots,a_{2^k-1})$ be the values, with $a_i\in[v_{\min},v_{\max}]$. QPIXL maps each value linearly to an angle $\theta_i=\pi\,(a_i-v_{\min})/(v_{\max}-v_{\min})\in[0,\pi]$. Hadamard gates prepare the address qubits in $\ket{+}^{\otimes k}$. A multiplexer then rotates the data qubit by $\RY(\theta_i)$ when the address qubits are in the computational basis state $\ket{i}$, where the $k$ bits of $i$ are the states of the $k$ address qubits. The result is the state $2^{-k/2}\sum_i\ket{i}\otimes\RY(\theta_i)\ket{0}$: conditioned on the address state $\ket{i}$, the data qubit is in
\begin{equation}
  \RY(\theta_i)\ket{0}=\cos\tfrac{\theta_i}{2}\ket{0}+\sin\tfrac{\theta_i}{2}\ket{1}.
  \label{eq:state}
\end{equation}
A measurement of the data qubit therefore gives $1$ with probability $P(1\mid i)=\sin^2(\theta_i/2)$ when the address qubits give $i$. 

\begin{figure}[t]
\centering
\scalebox{0.8}{\begin{quantikz}[row sep={0.6cm,between origins}, column sep=0.15cm]
\lstick{$a_0$} & \gate{H} & & \ctrl{2} & & & & \ctrl{2} & & & \\
\lstick{$a_1$} & \gate{H} & & & & \ctrl{1} & & & & \ctrl{1} & \\
\lstick{$d$}   & & \gate{\alpha_0} & \targ{} & \gate{\alpha_1} & \targ{} & \gate{\alpha_2} & \targ{} & \gate{\alpha_3} & \targ{} &
\end{quantikz}}\\[-2pt]
{\small (a)}\\[4pt]
\scalebox{0.8}{\begin{quantikz}[row sep={0.6cm,between origins}, column sep=0.15cm]
\lstick{$a_0$} & \gate{H} & & \ctrl{1} & & & & \ctrl{1} & & & \\
\lstick{$d_1$} & & \gate{\alpha^{(1)}_0} & \targ{} & \gate{\alpha^{(1)}_1} & \targ{} & \gate{\alpha^{(1)}_2} & \targ{} & \gate{\alpha^{(1)}_3} & \targ{} & \\
\lstick{$a_1$} & \gate{H} & & \ctrl{1} & & \ctrl{-1} & & \ctrl{1} & & \ctrl{-1} & \\
\lstick{$d_2$} & & \gate{\alpha^{(2)}_0} & \targ{} & \gate{\alpha^{(2)}_1} & \targ{} & \gate{\alpha^{(2)}_2} & \targ{} & \gate{\alpha^{(2)}_3} & \targ{} & \\
\lstick{$a_2$} & \gate{H} & & & & \ctrl{-1} & & & & \ctrl{-1} &
\end{quantikz}}\\[-2pt]
{\small (b)}
\caption{Multiplexer circuits. A box labelled $\alpha$ is the gate $\RY(\alpha)$. (a) One QPIXL multiplexer with $k=2$ address qubits: at step $j$ the data qubit $d$ receives $\RY(\alpha_j)$, then a CNOT from the address qubit whose bit flips in the $2$-bit Gray code $00,01,11,10$. (b) The smallest Tessa circuit, on a row of five qubits. The data qubits $d_1$ and $d_2$ share the address qubit $a_1$, while $a_0$ and $a_2$ each control one data qubit. In QCrank all addresses would be used by each data qubit. The two multiplexers run in lock-step: at every step their CNOTs act on disjoint pairs of qubits. }
\label{fig:circuits}
\end{figure}
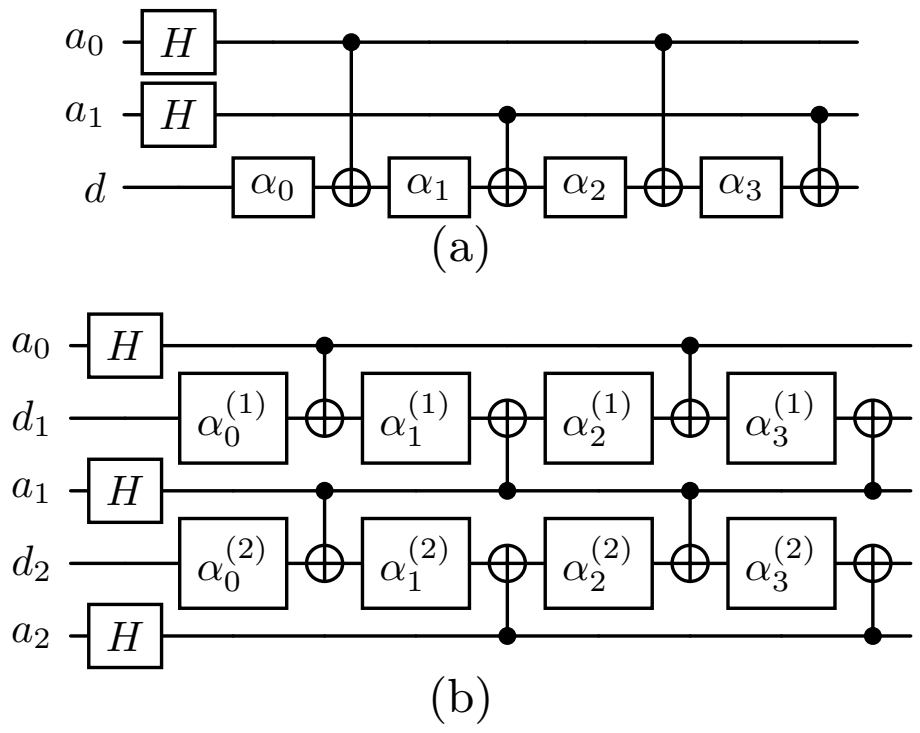

The multiplexer compiles into $2^k$ $\RY$ gates on the data qubit, alternated with $2^k$ CNOT gates, each controlled by one address qubit and targeting the data qubit~\cite{mottonen2005,shende2006,qpixl}. Let $g_0,\dots,g_{2^k-1}$ be the $k$-bit reflected Gray code, a sequence of all $k$-bit strings in which consecutive strings differ in exactly one bit, with $g_0=0$. At step $j$, the circuit applies $\RY(\alpha_j)$ to the data qubit, then a CNOT from the address qubit whose bit differs between $g_j$ and $g_{j+1}$ (indices modulo $2^k$). Before step $j$, the CNOTs have applied $X$ to the data qubit a number of times with parity $i\cdot g_j$, the bitwise inner product modulo $2$. Since $X\RY(\alpha)X=\RY(-\alpha)$ and the last CNOT returns the parity to zero, the total rotation for the address state $\ket{i}$ is $\theta_i=\sum_j(-1)^{i\cdot g_j}\alpha_j$. Inverting this relation gives
\begin{equation}
  \alpha_j=2^{-k}\sum_{i=0}^{2^k-1}(-1)^{i\cdot g_j}\,\theta_i .
  \label{eq:alpha}
\end{equation}
These two steps are called the scaled fast Walsh--Hadamard transform (sfwht) and the Gray-code permutation. Figure~\ref{fig:circuits}(a) shows the circuit for $k=2$. Setting to zero the $\alpha_j$ whose magnitude is below a threshold removes those $\RY$ gates, and adjacent CNOTs then cancel in pairs. This is the lossy compression of QPIXL~\cite{qpixl}. This applies without change to every multiplexer in Tessa.

QCrank~\cite{qcrank} adds data qubits that share one address register. Each data qubit is the target of its own multiplexer with the same controls. The multiplexers commute, because each is diagonal on the address qubits and targets a different qubit.

\subsection{Tessa}
\label{sec:pair}
Tessa differs from QCrank in one respect: each data qubit is controlled only by its neighbours on the device. These neighbourhoods overlap, so QCrank-like blocks tessellate the device, in the way that QCrank extends QPIXL.

Let the coupling graph $G=(V,E)$ be bipartite, $V=\dq\cup\aq$, with every edge joining a data qubit to an address qubit. If the device graph is not bipartite, removing the couplers between qubits of the same colour gives a bipartite subgraph with a lower degree. On a grid, the qubit in row $r$ and column $s$ is a data qubit if $r+s$ is even (Fig.~\ref{fig:lattice}). For $d\in\dq$, let $N(d)\subseteq\aq$ be its address neighbours and $m_d=|N(d)|$ its degree. We label the computational basis states of $N(d)$, the local address configurations, by integers $i\in\{0,\dots,2^{m_d}-1\}$. A data qubit and its neighbours form a super-pixel, which stores one value $a^{(d)}_i$, mapped to an angle $\theta^{(d)}_i$ as in Sec.~\ref{sec:background}, in each slot $(d,i)$. The capacity is the number of slots,
\begin{equation}
  \mathcal{C}=\sum_{d\in\dq} 2^{m_d}.
  \label{eq:capacity}
\end{equation}

The encoding unitary is
\begin{align}
  U&=\Big(\textstyle\prod_{d\in\dq}U_d\Big)\Big(\textstyle\bigotimes_{a\in\aq}H_a\Big),\nonumber\\
  U_d&=\sum_{i=0}^{2^{m_d}-1}\ketbra{i}{i}_{N(d)}\otimes \RY\!\big(\theta^{(d)}_i\big)_d ,
  \label{eq:U}
\end{align}
where a subscript names the qubits an operator acts on, and every operator acts as the identity elsewhere. Each $U_d$ is the multiplexer of Sec.~\ref{sec:background} with target $d$ and controls $N(d)$. The $U_d$ commute, because each is diagonal on the address qubits and acts on its own data qubit, so their order changes only the depth of the compiled circuit. App.~\ref{sec:depth} gives a schedule of minimal depth. Applying $U$ to $\ket{0}^{\otimes|V|}$ gives
\begin{equation}
  \ket{\Psi}
  =\frac{1}{2^{|\aq|/2}}\sum_{x\in\{0,1\}^{|\aq|}}\ket{x}_{\aq}
   \otimes\bigotimes_{d\in\dq}\ket{\psi^{(d)}_{x_{N(d)}}}_d ,
  \label{eq:fullstate}
\end{equation}
where $\ket{\psi^{(d)}_{i}}=\RY\big(\theta^{(d)}_{i}\big)\ket{0}$ and $x_{N(d)}$ is the configuration of $N(d)$ within the address string $x$. For each $x$, every data qubit receives one rotation, selected by its neighbours. The state is a superposition of product states of the data qubits, locally entangled with the address register (App.~\ref{app:entanglement}).

\paragraph{A second value per slot.}
The encoding above leaves the phase of each conditional state unused. If we replace $\RY(\theta^{(d)}_i)$ in Eq.~\eqref{eq:U} by $\RZ(\varphi^{(d)}_i)\RY(\theta^{(d)}_i)$, each slot holds a pair of angles, and the conditional state becomes
\begin{equation}
  \ket{\psi_i}=e^{-\mathrm{i}\varphi_i/2}\Big(\cos\tfrac{\theta_i}{2}\ket0+e^{\mathrm{i}\varphi_i}\sin\tfrac{\theta_i}{2}\ket1\Big),
  \label{eq:pairstate}
\end{equation}
with the superscript $(d)$ dropped and $\mathrm{i}$ the imaginary unit. The factor $e^{-\mathrm{i}\varphi_i/2}$ depends only on the branch $i$. The $\RZ$ multiplexer compiles like the $\RY$ one and adds one CNOT per value. It does not commute with the $\RY$ multiplexer on the same target, so we apply all $\RY$ multiplexers and then all $\RZ$ multiplexers, which doubles the depth. The azimuth is undefined at the poles of the Bloch sphere, so we restrict the polar angle to $\theta\in[c,\pi-c]$, where $c=0.2\pi$ gives the lowest error (App.~\ref{sec:clamp}).

\begin{figure*}[t]
\centering
\begin{tabular}{@{}l@{\hspace{2pt}}c@{\hspace{8pt}}l@{\hspace{2pt}}c@{}}
\raisebox{-\height}{\small (a)} &
\raisebox{-\height}{\includegraphics[width=0.46\linewidth]{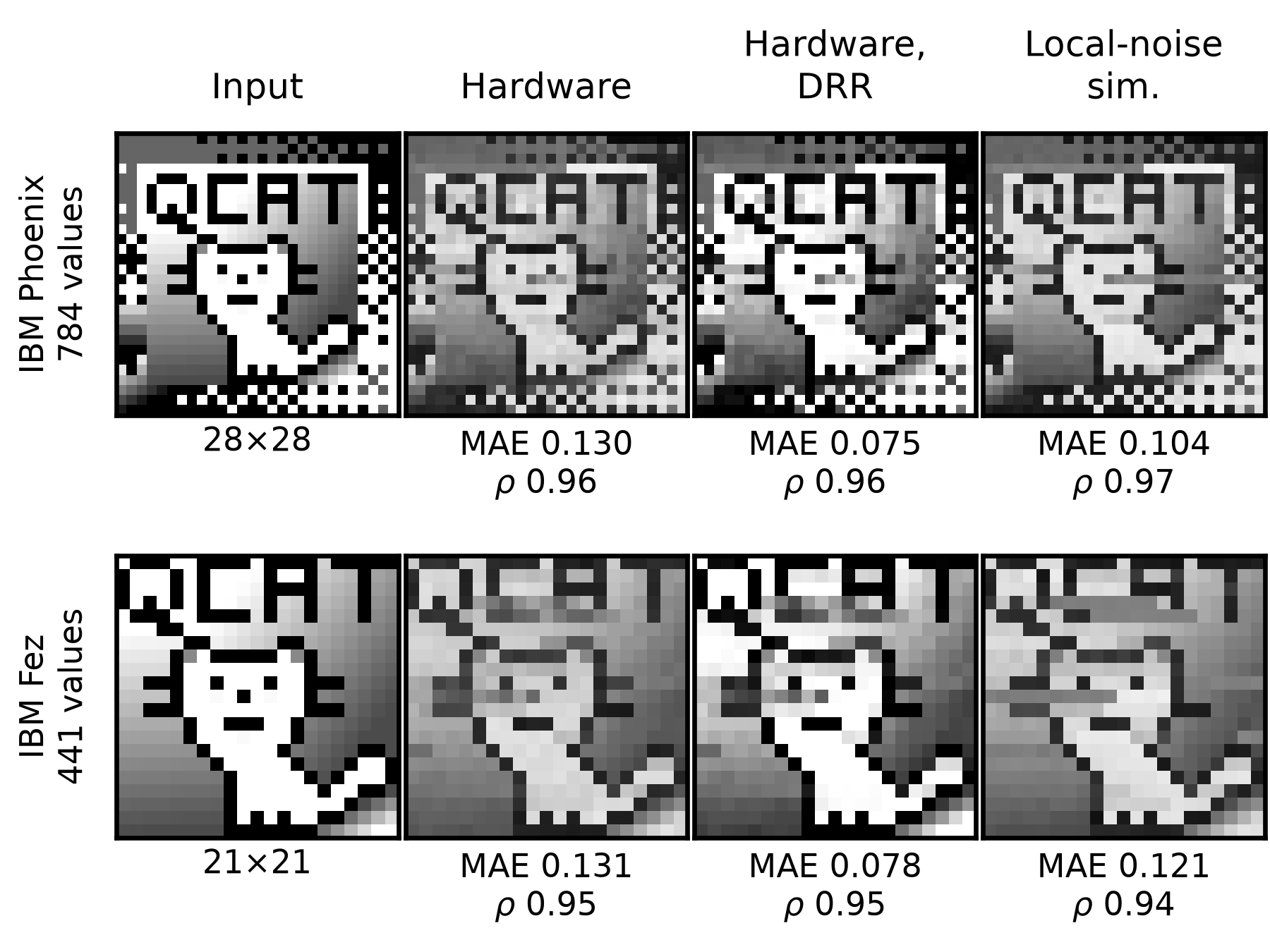}} &
\raisebox{-\height}{\small (b)} &
\raisebox{-\height}{\includegraphics[width=0.46\linewidth]{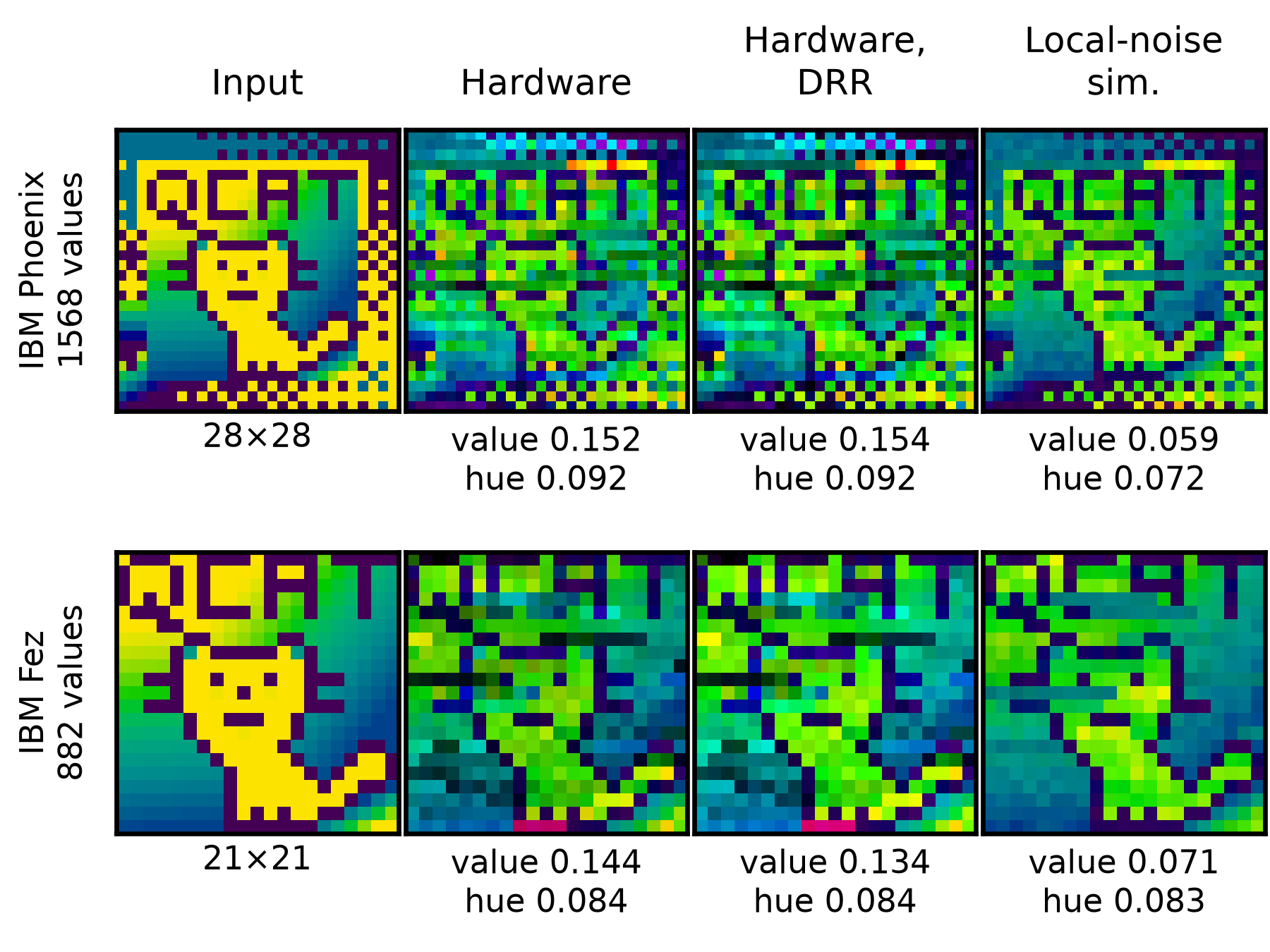}}
\end{tabular}
\caption{Whole-chip QCAT runs at $8192$ shots per basis in one job per device. In each panel, the top row is a $28\times28$ image on Phoenix and the bottom row a $21\times21$ image on Fez. From the left: the input, the raw hardware reconstruction, the same after \texttt{mean\_std} dynamic-range recovery (DRR), and the local-noise simulation of the same circuits.
(a) Single basis encoding, $Z$ readout. Below each panel are the mean absolute error and the correlation $\rho$ with the input.
(b) Pair encoding, $Z,X$ readout: the value in $\theta$ and the hue in $\varphi$, rendered at full saturation. Below each panel are the value and hue errors.}
\label{fig:qcat}
\end{figure*}

\subsection{Readout}
\label{sec:independence}
Data qubits that share an address neighbour are correlated through it. The following proposition shows that we can still read their data separately.

\begin{proposition}
\label{prop:indep}
For a data qubit $d$, let $S=\{d\}\cup N(d)$ be its super-pixel and $\rho_S=\operatorname{Tr}_{\overline{S}}\ketbra{\Psi}{\Psi}$ its reduced state, where $\overline S$ is the set of all other qubits. Then, for every local address configuration $i$,
\begin{equation*}
  \big(\bra{i}_{N(d)}\otimes\mathbb{I}_d\big)\,\rho_S\,\big(\ket{i}_{N(d)}\otimes\mathbb{I}_d\big)
  =2^{-m_d}\,\ketbra{\psi^{(d)}_i}{\psi^{(d)}_i},
\end{equation*}
with $\ket{\psi^{(d)}_i}$ as in Eq.~\eqref{eq:fullstate} or \eqref{eq:pairstate}. This block does not depend on the angles of other data qubits.
\end{proposition}
\noindent
The proof is in App.~\ref{app:proof}. It holds unchanged for the pair encoding.

We measure every qubit in the computational ($Z$) basis. By Eq.~\eqref{eq:fullstate}, once the address outcome $x$ is fixed, the data outcomes $y_d$ are independent Bernoulli variables with $P(y_d=1\mid x)=\sin^2\!\big(\theta^{(d)}_{x_{N(d)}}/2\big)$. For each data qubit $d$ we keep only the outcomes of its super-pixel and count the shots $n^{(d)}(i,y)$ with configuration $i$ and data outcome $y$. With $P^{(d)}_y(i)=n^{(d)}(i,y)/\big(n^{(d)}(i,0)+n^{(d)}(i,1)\big)$, we invert
\begin{equation}
  \theta^{(d)}_i = 2\operatorname{atan2}\!\Big(\sqrt{P^{(d)}_1(i)},\,\sqrt{P^{(d)}_0(i)}\Big).
  \label{eq:decode}
\end{equation}
With exact probabilities this recovers every angle exactly, and one set of device-wide shots gives the marginals of every data qubit. The phase is read in the same way after rotating only the data qubits into the $X$ or $Y$ basis (App.~\ref{sec:xyz}). This readout procedure does not recover off-diagonal elements, but they are recoverable and meaningful (App.~\ref{sec:downstream}).

The cost of the readout is in the shot count. Each of the $2^{m_d}$ local address configurations of a data qubit occurs in a fraction $2^{-m_d}$ of the $N$ shots, so each value is estimated from about $n=N/2^{m_d}$ shots. To leading order in $1/n$, $\operatorname{Var}(\hat\theta^{(d)}_i)\approx1/n$ (App.~\ref{app:proof}), so the error scales as $2^{m_d/2}/\sqrt{N}$. The factor $2^{m_d}$ that lets a data qubit of degree four store $16$ values also divides its shots among them.

\section{Hardware experiments}\label{sec:hardware}
\begin{figure}[!b]
\centering
\begin{tabular}{cc}
\includegraphics[width=\columnwidth/3]{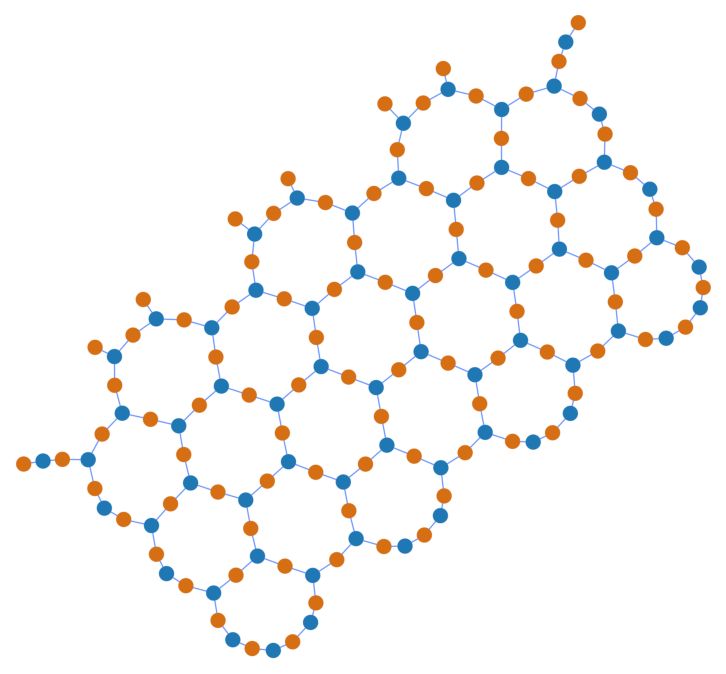} &
\includegraphics[width=\columnwidth*3/8]{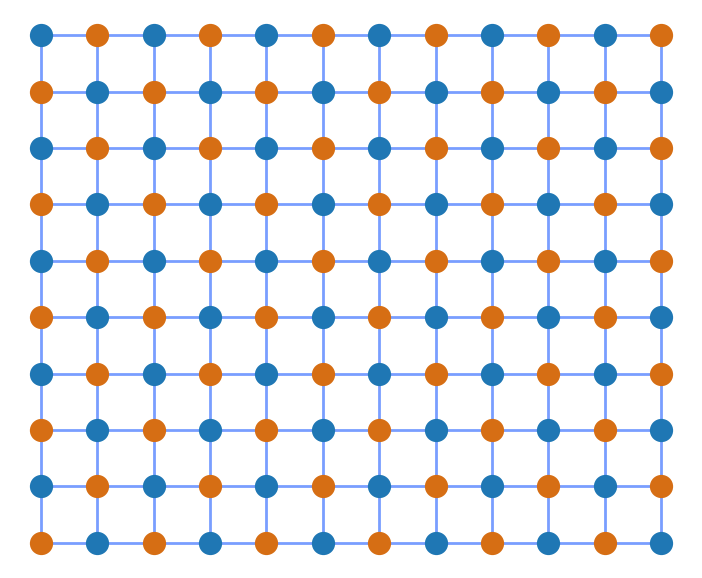}\\
(a) heavy hex & (b) square lattice
\end{tabular}
\caption{Hardware coupling graphs with their maximum-capacity 2-colouring (data in blue, address in orange) as used in the experiments.}
\label{fig:hh}
\end{figure}
We can 2-colour any bipartite coupling graph into data and address qubits. For each connected component we choose as data qubits the colour class that maximises the capacity. This maximal capacity colouring is shown in Fig.~\ref{fig:hh}(a) for heavy hex devices like IBM's Fez ($\mathcal C=448$) and Fig.~\ref{fig:hh}(b) for square lattice devices like Phoenix ($\mathcal C=792$). 

No classical computer can hold a $120+$ qubit statevector, but Proposition~\ref{prop:indep} allows a simple circumvention of the problem. In the noiseless circuit, the readout statistics of a super-pixel equal those of its own sub-circuit, whatever the angles on the rest of the chip (App.~\ref{app:proof}). We can therefore simulate each super-pixel separately.

We also use this per-super-pixel simulation, with noise models derived from the calibration data. It is only a rough approximation of the device. It omits errors that couple a super-pixel to qubits outside it, such as crosstalk. A more precise simulation would include the neighbouring qubits of each super-pixel at minimum, but is out of scope for this work. 

\subsection{Single basis encoding}
\label{sec:hwrun}

To show the near-term performance of Tessa, we encoded a pixel-art image of a cat called QCAT, on both devices, and recovered it using a computational basis readout in one job of $8192$ shots (Fig.~\ref{fig:qcat}a, Table~\ref{tab:hw}). On Phoenix a $28\times28$ image fills $784$ of the $792$ slots. On Fez the $21\times21$ image fills $441$ of $448$ slots. The spare slots held zeros, so the experiment used every qubit. For each run we also simulate the same circuits at the same shot count: once with a noise model derived from that run's calibration, and once without noise.

The raw hardware output shows the image. The letters, the cat and, on Phoenix, the test-pattern border are legible. Most of the error is a compression of the range towards its centre. Dynamic-range recovery (DRR, App.~\ref{sec:drr}) corrects for this with a few summary statistics of the input. In its simplest form, it stretches the reconstruction so that its minimum and maximum equal those of the input. DRR cuts the mean error by about two-fifths. The local-noise simulation predicts the mean error to within $0.03$ (Table~\ref{tab:hw}), but its images differ from the hardware images. In the simulation, the faulty qubits and couplers are visually worse than on the real device. 

The main result is that retrieval works at full-chip scale on two different coupling graphs. It also shows that the measured diagonal blocks of the prepared state are close to those expected. We use no error mitigation in this work, because our aim is to test the encoding itself. Error mitigation techniques should be able to reduce the reconstruction error significantly.

\begin{figure*}[p]
\centering
\includegraphics[width=\linewidth]{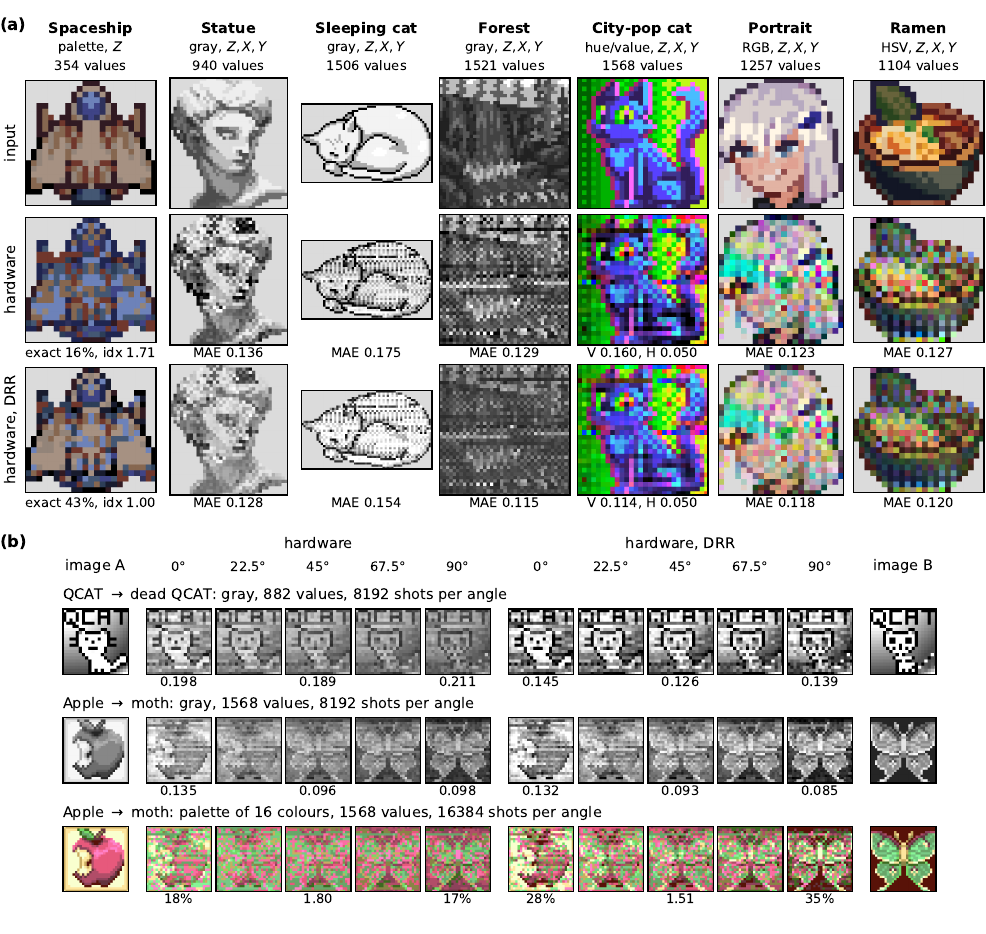}
\caption{Pixel art on Phoenix with $8192$ shots per circuit unless noted. Transparent pixels are not encoded and are shown in light gray.
(a) One image per column: the input, the raw hardware reconstruction, and the reconstruction after the dynamic-range recovery (DRR, App.~\ref{sec:drr}) that best matches the input (\texttt{mean\_std} for Spaceship, Sleeping cat and City-pop cat; \texttt{percentile} for Statue and Portrait; \texttt{min\_avg\_max} for Forest and Ramen). The header gives the representation, the measurement bases and the number of real values stored.
\emph{Palette}: each pixel stores its index in the palette of the image ($13$ colours, sorted dark to light), one per slot, read in $Z$ and rounded to the nearest index. Below the panels are the fraction of exact pixels and the mean index error.
\emph{Gray}: the luminance of each pixel, two pixels per slot, one in $\theta$ and one in $\varphi$ on the full circle, read in $Z$, $X$ and $Y$ (App.~\ref{sec:xyz}).
\emph{Hue/value}: the value in $\theta$ and the hue in $\varphi$. The saturation is taken from the input and is not encoded.
\emph{RGB} and \emph{HSV}: three channels per pixel, two numbers per slot.
Errors are mean absolute errors on $[0,1]$: of the luminance for gray, of the rendered RGB image for RGB and HSV, and of the value (V) and the hue as a fraction of the circle (H) for hue/value.
(b) Transitions. Image A is stored in $\theta$ and image B in $\varphi$ on the half circle (Sec.~\ref{sec:pair}). Before the measurement, each data qubit is rotated so that a $Z$ measurement reads $\cos\alpha\,Z+\sin\alpha\,X$. The angle $\alpha=0$ returns A, $\alpha=90^\circ$ returns B, and the angles between give a blend of the two. One circuit per angle. Each row shows image A, the raw frames, the frames after \texttt{mean\_std} DRR, and image B. Below the frames at $0^\circ$, $45^\circ$ and $90^\circ$ are the mean absolute error against A at $\alpha=0$, the mean over all five angles against the noiseless frames, and the error against B at $\alpha=90^\circ$. The palette transition uses one palette of $16$ colours for both images, and its numbers are the percentage of exact pixels at $0^\circ$ and $90^\circ$, under $45^\circ$ is the MAE of the index value. More details in App.~\ref{app:music}.
Pixel art: Spaceship, Statue, Portrait and Ramen by M.A.~Kozhikkal (SalvaPixel); City-pop cat, Apple and Moth by M.~Fish (Leviatic); Forest by D.~Souza; Sleeping cat by E.~K{\i}l{\i}n\c{c} (Diaspixels); QCAT by the author; dead QCAT by S.~Cantori.}
\label{fig:pixelart}
\end{figure*}

\subsection{Pair encoding}
\label{sec:pairsim}

The pair encoding was tested on a coloured version of the QCAT image, since a fully saturated colour can be represented with two numbers: value and hue.
By rendering the QCAT image in the viridis colour map and storing the value in $\theta$ and the hue in $\varphi$, we ran the pair encoding with $Z,X$ readout on both devices: the $21\times21$ colour image on Fez, with $882$ values, and the $28\times28$ colour image on Phoenix, with $1568$ values (Fig.~\ref{fig:qcat}b, Table~\ref{tab:hw}). The remaining slots held zeros.

The pair encoding is less accurate than the single encoding. The correlation of the value with the input falls from $0.95$--$0.96$ to $0.69$--$0.73$ (Table~\ref{tab:hw}), and DRR no longer reduces the value error significantly, so the extra error is not only a compression of the range, which is expected because there is no amplitude damping. On hardware, the value error is two to three times that of the local-noise simulation, while the hue error is close to it on Fez ($0.084$ against $0.083$). The noiseless error is also higher than for the single encoding ($0.013$--$0.018$ against $0.004$), because the clamp $c=0.2\pi$ shrinks the range of $\theta$ (App.~\ref{sec:clamp}). The pair circuits have twice as many CZ gates, and the larger gap between hardware and simulation shows the limits of the approximate noise model. Many more examples of different data and encoding strategies can be found in Fig.~\ref{fig:pixelart} and App.~\ref{app:music}.

\begin{table}[t]
\centering
\caption{Hardware runs with $8192$ shots per basis, each paired with a noisy simulation using that
machine's calibration data (``local sim.'') and noiseless simulation. Values come from computational basis measurements; in the
pair runs of Sec.~\ref{sec:pairsim}, hues come from $X$ (clamp $c=0.2\pi$). MAE is the
mean absolute error, before and after \texttt{mean\_std} recovery (DRR). $\rho$ is the
correlation of the value with the input.}
\label{tab:hw}
\footnotesize\setlength{\tabcolsep}{4pt}
\begin{tabular}{lcccc}
\toprule
source & value MAE & DRR & value $\rho$ & hue MAE\\
\midrule
\multicolumn{5}{l}{Phoenix, $Z$ ($784$ values, $728$ CZ)}\\
\quad hardware   & 0.130 & 0.075 & 0.96 & ---\\
\quad local sim. & 0.104 & 0.059 & 0.97 & ---\\
\quad noiseless  & 0.004 & 0.004 & 1.00 & ---\\
\multicolumn{5}{l}{Fez, $Z$ ($441$ values, $428$ CZ)}\\
\quad hardware   & 0.131 & 0.078 & 0.95 & ---\\
\quad local sim. & 0.121 & 0.081 & 0.94 & ---\\
\quad noiseless  & 0.004 & 0.004 & 1.00 & ---\\
\midrule
\multicolumn{5}{l}{Phoenix, $Z,X$ ($1568$ values, $1456$ CZ)}\\
\quad hardware   & 0.152 & 0.154 & 0.69 & 0.092\\
\quad local sim. & 0.059 & 0.048 & 0.94 & 0.072\\
\quad noiseless  & 0.018 & 0.017 & 1.00 & 0.025\\
\multicolumn{5}{l}{Fez, $Z,X$ ($882$ values, $852$ CZ)}\\
\quad hardware   & 0.144 & 0.134 & 0.73 & 0.084\\
\quad local sim. & 0.071 & 0.057 & 0.93 & 0.083\\
\quad noiseless  & 0.013 & 0.012 & 1.00 & 0.015\\
\bottomrule
\end{tabular}
\end{table}

\subsection{Comparison with earlier experiments}
We compare Tessa only with experiments that wrote a list of arbitrary real numbers into a physical quantum processor and read them back.
 This excludes experiments that encode data but read back only one derived quantity. For example, in~\cite{bluequbit2025} representations of two-million-pixel images were loaded onto $72$ qubits and classified.

By the preceding definition we only compare \emph{how many} values are recovered. This is because the quality of a reconstruction depends on the device, on its calibration that day and on the metric chosen, which is not consistent.
Table~\ref{tab:prior} lists the experiments by count. QCrank recovered $128$ values from twelve qubits of a trapped-ion machine~\cite{qcrank}. A later compilation study reports configurations of six to twenty qubits carrying up to $320$ values; it ran the smaller ones on Fez and simulated the largest~\cite{qcrankdpqa}. A recent study recovered $1024$ pixels from thirteen qubits of Fez with a trained tensor-network approximation~\cite{tnqe}. That study encodes a unitary tensor-network approximation of the data, not an arbitrary list of reals, but it is a promising route.

Amplitude encoding stores $2^n$ values on $n$ qubits in principle, so Tessa doesn't come close to saturating this limit. Other encodings can in theory store many more values, but the transpiled circuits need to account for routing which limits the amount of data that can be encoded and decoded. The pair encoding experiment with up to $1568$ values from $120$ qubits encodes and decodes more values than any earlier experiment under this definition. This is a higher encoding capacity than a product state of QPIXL blocks on the chip would achieve. 
\begin{table}[t]
\centering
\caption{Experiments that encoded and decoded data on hardware with post-transpilation depths. Missing information is represented by a dash. Two-qubit (2Q) gate counts and depths are after transpilation to the device.}
\label{tab:prior}
\footnotesize\setlength{\tabcolsep}{2pt}
\resizebox{\linewidth}{!}{%
\begin{tabular}{llcccc}
\toprule
work & device & qubits & values & 2Q & 2Q\\
 & & & & gates & depth\\
\midrule
QCrank~\cite{qcrank}                 & trapped ion             & $12$      & $128$      & $128^{a}$    & $128^{a}$\\
QCrank~\cite{qcrankdpqa} & Fez       & $6$--$12$ & $\le 72$   & $66$--$262$  & $\sim\!45$--$190$\\
TNQE~\cite{tnqe}                     & Fez       & $13$      & $1024^{b}$ & ---          & ---\\
Tessa, $Z$ (ours)                    & Phoenix   & $120$     & $784$      & $728^{c}$        & $23$\\
Tessa, $Z,X$ (ours)                  & Phoenix   & $120$     & $1568$     & $1456^{c}$       & $43$\\
\bottomrule
\end{tabular}}\\[2pt]
\parbox{\linewidth}{\scriptsize
$^{a}$estimated from a transpilation of the stated QCrank circuits.\\
$^{b}$a tensor-network approximation, not an arbitrary list of values.\\
$^{c}$when an angle is 0, the CNOT isn't necessarily applied.}
\end{table}

\section{Conclusion}
\label{sec:conclusion}
We have presented Tessa, a lattice form of QCrank with the address register embedded in the device coupling graph. Sharing address qubits raises the storage density while allowing the data to remain recoverable. 
The method needs only a bipartite coupling graph and can be made to run without any routing, but it is not limited to that, hence one can greatly increase the encoding capacity with a controllable cost in depth and shots. Reading the payload has a shot cost that grows exponentially with the local degree, but this can be kept small, as we did here. Using Tessa, our hardware experiments returned recognisable images from every qubit of two chips, with errors close to our local noise model.

\subsubsection*{Acknowledgements}
I thank J.S. Ferreira, S. Topel, E. Markevičiūtė and S. Cantori for useful discussions, and J. Wootton and P. Fromholz for comments on the manuscript. Moth Quantum supported this work. Additional non-QCAT pixel art was provided by M.A.~Kozhikkal (SalvaPixel), M.~Fish (Leviatic), D.~Souza and E.~K{\i}l{\i}n\c{c} (Diaspixels). 

We acknowledge the use of IBM Quantum Credits via the IBM Quantum Startups Program for this work. The views expressed are those of the authors and do not reflect the official policy or position of IBM or the IBM Quantum Platform team.

\subsubsection*{Code and Data Availability}
Code and data are available at in the repository: \href{https://github.com/moth-quantum/Tessa}{\texttt{https://github.com/moth-quantum/Tessa}}. All experiments shown where run on \texttt{ibm\_fez}, a 156-qubit IBM Quantum Heron r2 chip and \texttt{ibm\_phoenix}, a 120 qubit IBM Nighthawk r2 quantum processor. Code was written in part with Qiskit \cite{qiskit2024} and run ont the IBM quantum platform \cite{ibm_quantum_platform}.

\bibliographystyle{unsrtnat}
\bibliography{refs}
\clearpage
\appendix
\section{Optimal circuit construction}
\label{sec:depth}

Each multiplexer is a chain of $2^{m_d}$ steps, each one $\RY$ gate and one CNOT on its data qubit (Sec.~\ref{sec:background}). Applied in sequence, the chains give a depth that grows linearly with $|\dq|$. Since the multiplexers commute, we interleave them, so that step $j$ of every chain takes place in the same layers (Fig.~\ref{fig:circuits}(b)). This requires that no two data qubits use the same address qubit in one step.

A proper edge colouring $\chi:E\to\{0,\dots,m-1\}$ gives different colours to edges that share a qubit, so each colour class is a matching, a set of disjoint edges. By K\"onig's theorem~\cite{konig1916}, a bipartite graph of maximum degree $m$ has one with $m$ colours. Suppose every data qubit has degree $m$ and no address qubit has a larger degree. We let bit $b$ of the local address configuration of $d$ be the neighbour joined to $d$ by the edge of colour $b$. At step $j$ every data qubit then flips the same bit $b_j$, all CNOTs lie on the matching of colour $b_j$, and the step takes one $\RY$ layer and one CNOT layer. The interleaved depth is $2\cdot2^{m}$, that of a single multiplexer, for any number of data qubits. This is optimal, since each data qubit is the target of $2^m$ sequential CNOTs. On the square lattice the colouring is by direction (Fig.~\ref{fig:lattice}).

Boundary data qubits have a lower degree and shorter chains, so the chains no longer flip the same bit in the same step. The edges used in one step give each address qubit degree at most $\Delta_{\aq}$ (the largest number of data qubits adjacent to one address qubit). This splits them into at most $\Delta_{\aq}$ matchings. The interleaved depth is then at most $(1+\Delta_{\aq})\,2^{m_{\max}}$, where $m_{\max}$ is the largest degree of a data qubit. This bound is loose but independent of the lattice size. On IBM's native gates $\{\sqrt{X},\RZ,\mathrm{CZ}\}$, each CNOT becomes a CZ whose single-qubit gates merge with the adjacent $\RY$ gates, so compilation keeps the number of CZ layers and raises only the single-qubit gate depth.

\section{Proof of Proposition~\ref{prop:indep}}
\label{app:proof}

Let $G=(\dq\cup\aq,E)$ be the bipartite coupling graph. For a data qubit $d$, $N(d)$ is the set of its address neighbours, $m_d=|N(d)|$, and $\ket{\psi^{(d)}_i}=\RY(\theta^{(d)}_i)\ket0$ for $i=0,\dots,2^{m_d}-1$. For an address string $x\in\{0,1\}^{|\aq|}$ and a set $B\subseteq\aq$, $x_B$ is the configuration of the qubits of $B$ within $x$. The encoded state of Eq.~\eqref{eq:fullstate} is
\begin{equation}
  \ket{\Psi}=\frac{1}{2^{|\aq|/2}}\sum_{x\in\{0,1\}^{|\aq|}}\ket{x}_{\aq}
  \bigotimes_{d'\in\dq}\ket{\psi^{(d')}_{x_{N(d')}}}_{d'} .
  \label{eq:app:psi}
\end{equation}
Here we use the bipartiteness of $G$. No data qubit is adjacent to another, so no data qubit acts as a control. The state of the data qubits, given the address string $x$, is therefore an exact product over $\dq$, and the factor of each $d'$ depends on $x$ only through $x_{N(d')}$. If two data qubits were adjacent, the conditional state would not factorise and the argument would fail.

Fix the data qubit $d$ and let $S=\{d\}\cup N(d)$. Split the address qubits as $\aq=N(d)\,\dot\cup\,\aq_o$ and the data qubits as $\dq=\{d\}\,\dot\cup\,\dq_o$, so that the traced-out set is $\overline S=\aq_o\cup\dq_o$. Write an address string as $x=(u,v)$, with $u=x_{N(d)}$ and $v=x_{\aq_o}$. We abbreviate the state of the other data qubits as
\begin{equation}
  \ket{\Xi_{u,v}}=\bigotimes_{d'\in\dq_o}\ket{\psi^{(d')}_{(u,v)_{N(d')}}} ,
  \qquad \braket{\Xi_{u,v}}{\Xi_{u,v}}=1 .
\end{equation}
Then $\ket{\Psi}=2^{-|\aq|/2}\sum_{u,v}\ket{u}_{N(d)}\ket{v}_{\aq_o}\ket{\psi^{(d)}_u}_d\ket{\Xi_{u,v}}_{\dq_o}$. The factor $\ket{\psi^{(d)}_u}$ depends on $u$ alone, because $N(d)$ is exactly the set of controls of $d$.

Expanding the projector,
\begin{multline}
  \ketbra{\Psi}{\Psi}=\frac{1}{2^{|\aq|}}\sum_{u,v}\sum_{u',v'}
  \ketbra{u}{u'}\otimes\ketbra{v}{v'}\\
  \otimes\;\ketbra{\psi^{(d)}_u}{\psi^{(d)}_{u'}}\otimes\ketbra{\Xi_{u,v}}{\Xi_{u',v'}} .
\end{multline}
The trace over $\aq_o$ gives $\delta_{v,v'}$. The trace over $\dq_o$ gives the scalar $\braket{\Xi_{u',v}}{\Xi_{u,v}}$. Hence
\begin{multline}
  \rho_S=\frac{1}{2^{|\aq|}}\sum_{u,u'}\left(\sum_{v}\braket{\Xi_{u',v}}{\Xi_{u,v}}\right)\\
  \times\;\ketbra{u}{u'}\otimes\ketbra{\psi^{(d)}_u}{\psi^{(d)}_{u'}} .
  \label{eq:app:rho}
\end{multline}
The bracketed scalar behaves differently on and off the diagonal.

\subsection*{Diagonal blocks}
Put $u=u'=i$. Each $\braket{\Xi_{i,v}}{\Xi_{i,v}}=1$, so the sum over $v$ counts $2^{|\aq_o|}$ terms. Since $|\aq|-|\aq_o|=m_d$, the block of $\rho_S$ at the local address configuration $i$ is
\begin{equation}
  2^{-m_d}\,\ketbra{\psi^{(d)}_i}{\psi^{(d)}_i} ,
\end{equation}
which is Proposition~\ref{prop:indep}. The other data qubits drop out identically, because each $\ket{\Xi_{i,v}}$ is a unit vector whatever their payloads.

\subsection*{Off-diagonal blocks}
For $u\neq u'$ the scalar does not collapse. Write $\Lambda_{u,u'}=2^{-|\aq_o|}\sum_v\braket{\Xi_{u',v}}{\Xi_{u,v}}$. Using $\braket{\psi^{(d')}_j}{\psi^{(d')}_k}=\cos\big((\theta^{(d')}_j-\theta^{(d')}_k)/2\big)$,
\begin{equation}
  \Lambda_{u,u'}=\mathop{\mathbb{E}}_{v}\prod_{d'\in\dq_o}
  \cos\!\left(\frac{\theta^{(d')}_{(u,v)_{N(d')}}-\theta^{(d')}_{(u',v)_{N(d')}}}{2}\right),
  \label{eq:lambda}
\end{equation}
with the expectation uniform over $v$. If $d'$ has no neighbour in $N(d)$, its cosine is $1$ and it drops out. Only data qubits at graph distance two from $d$ contribute. The off-diagonal blocks thus combine the conditional states of $d$ with a factor set by the angles of the neighbouring super-pixels. They are coherences of the same kind as the observable $K_a$ of App.~\ref{sec:downstream}: invisible to a computational-basis marginal, and visible after a change of basis on the address qubits.

\subsection*{Recovery of the angle}
Measuring $S$ in the computational basis gives $P(i,y)=\bra{i,y}\rho_S\ket{i,y}=2^{-m_d}|\braket{y}{\psi^{(d)}_i}|^2$ for the data outcome $y\in\{0,1\}$, since only the diagonal block contributes to a diagonal element. Hence $P(1\mid i)=\sin^2(\theta^{(d)}_i/2)$ and $P(0\mid i)=\cos^2(\theta^{(d)}_i/2)$. Because $\theta^{(d)}_i/2\in[0,\pi/2]$, where sine and cosine are both non-negative, the positive square roots are the correct branches, and Eq.~\eqref{eq:decode} inverts the encoding uniquely. The factor $\Lambda$ never appears, so the ordinary readout does not touch the correlations. \qed

\subsection*{Two corollaries}
The first is the whole device simulation approximation of Sec.~\ref{sec:hardware}. The off-diagonal blocks of $\rho_S$ depend on the neighbouring super-pixels through $\Lambda$, but the diagonal blocks depend on the payload only through $\theta^{(d)}$. The computational-basis statistics of $S$ use only the diagonal blocks, so they equal those of the sub-circuit on $S$ alone. A super-pixel has at most four qubits on heavy-hex, or five on a square lattice. One can therefore simulate each super-pixel alone, and the readout statistics hold for the whole noiseless device.

The second is the shot cost of Sec.~\ref{sec:independence}. The configuration $i$ occurs with probability $2^{-m_d}$, so about $n=N/2^{m_d}$ of the $N$ shots estimate $P(1\mid i)$, with variance $P(1-P)/n$. The delta method applied to $\theta=2\arcsin\sqrt{P}$, whose derivative is $1/\sqrt{P(1-P)}$, gives $\operatorname{Var}(\hat\theta)\approx 1/n=2^{m_d}/N$.

\section{Entanglement of the encoded state}
\label{app:entanglement}
The multiplexers commute, and every two-qubit gate is controlled from the address
register, but the encoded state is not a product state.
This follows from Eq.~\eqref{eq:fullstate}, and here we quantify it.
The Tessa state is entangled across the address--data cut, its entanglement is
short-ranged, and the amount is set by the local contrast of the payload.
All results in this appendix come from simulation.

\begin{figure}[t]
\centering
\includegraphics[width=\linewidth]{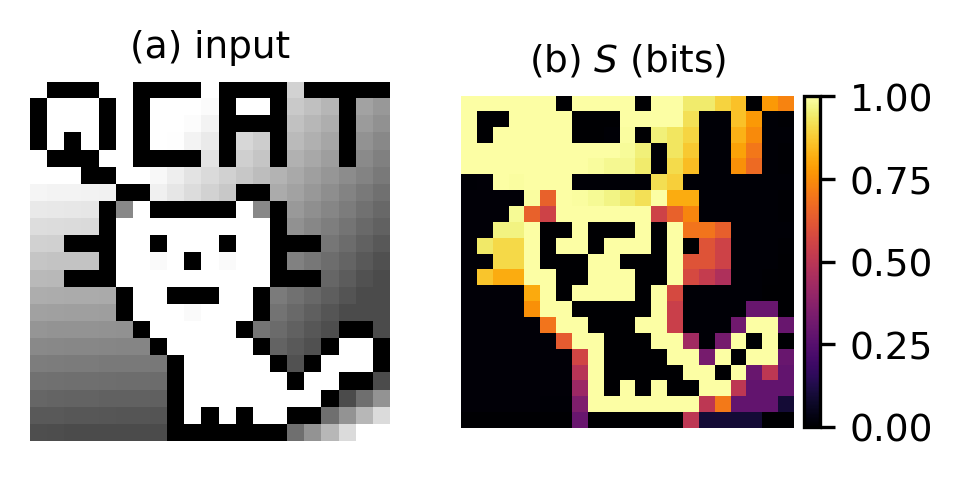}
\caption{Entanglement entropy $S=h\big((1+K)/2\big)$ for the address qubit of each
neighbouring-pixel pair of the QCAT image. Each
pixel shows the larger of its horizontal and vertical bonds. Flat regions carry no
entanglement; a hard edge carries one bit.}
\label{fig:entmap}
\end{figure}

Split the state on an address qubit $a$ as in App.~\ref{sec:downstream}. Tracing out
everything but $a$ gives
$\rho_a=\tfrac12\left(\begin{smallmatrix}1&K_a\\K_a&1\end{smallmatrix}\right)$, with
$K_a=\braket{\Phi_0}{\Phi_1}$ the coherence of Eq.~\eqref{eq:Ka}. The whole state is pure,
so the entanglement entropy of $a$ with the rest of the device is
\begin{equation}
  S_a=h\!\left(\frac{1+K_a}{2}\right),
  \label{eq:Sa}
\end{equation}
where $h$ is the binary entropy. $S_a=0$ exactly when $K_a=1$, that is, when flipping the
bit $a$ changes nothing. Otherwise $S_a>0$, reaching one bit at $K_a=0$. Any payload that
varies along some address direction therefore gives an entangled state. The interference
job of App.~\ref{sec:downstream} is thus also a measurement of the entanglement
entropy of every address qubit with the rest of the chip. Computing $S_a$ independently
from the reduced density matrix and from the marginal after the Hadamard layer gives the
same answer. Figure~\ref{fig:entmap} shows the entropy on the QCAT image. It sits on the
edges; flat regions carry none.

The entanglement is easy to remove by measuring the address register in the
computational basis. This leaves the data register in an exact product state as is done in the readout of
Sec.~\ref{sec:independence}. 

\subsection{Area law}
\begin{figure*}[t]
\centering
\includegraphics[width=0.85\linewidth]{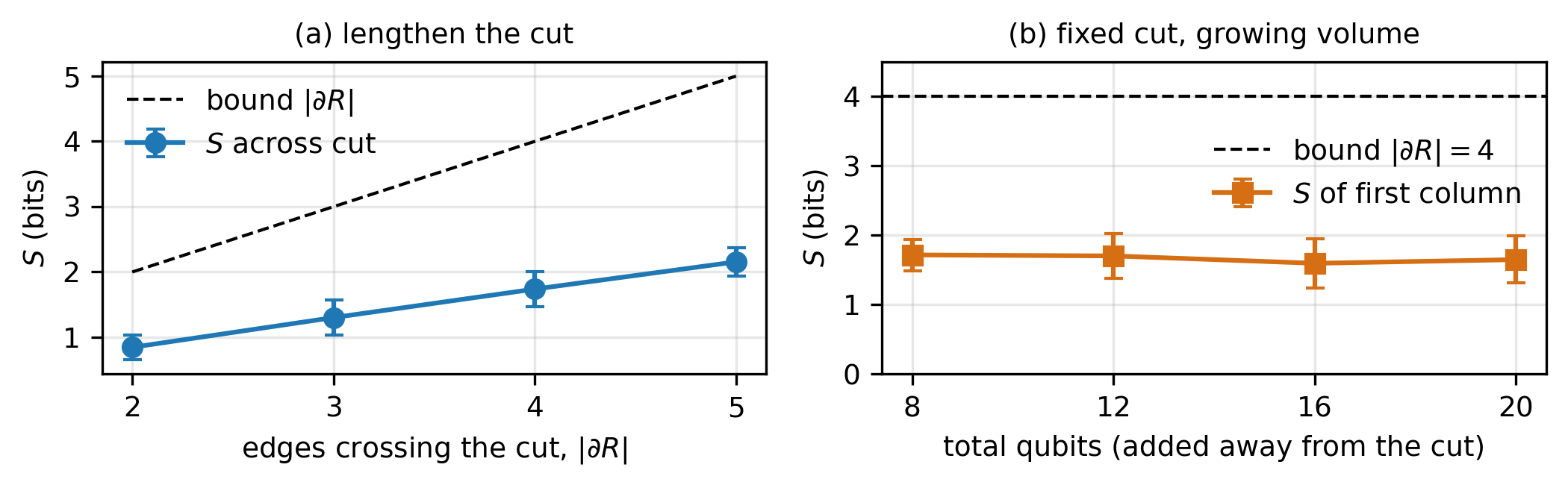}
\caption{Entanglement entropy across a cut, mean and standard deviation over $12$ random
payloads. (a) $R\times4$ lattices cut between the second and third columns, so that
$|\partial R|=R$ couplers are severed: $S$ grows linearly with the length of the cut and
stays below the bound of Eq.~\eqref{eq:arealaw}. (b) $4\times C$ lattices with the region
fixed as the first column, $|\partial R|=4$: adding columns away from the cut leaves $S$
unchanged within its spread.}
\label{fig:arealaw}
\end{figure*}

A tensor network gives an explicit bound. Attach to each address qubit a copy tensor,
with one physical leg and one virtual leg of dimension two per incident edge, which
broadcasts its bit to its data neighbours. Attach to each data qubit $d$ a tensor with
entries $\braket{y_d}{\psi^{(d)}_{x_{N(d)}}}$. Contracting the network reproduces
Eq.~\eqref{eq:fullstate} exactly, with bond dimension two on every edge of the coupling
graph. Hence, for any region $R$ with $|\partial R|$ couplers crossing its boundary,
\begin{equation}
  S(R)\le|\partial R|\ \text{bits}.
  \label{eq:arealaw}
\end{equation}
The entropy grows with the perimeter of a region and not with its area, as Fig.~\ref{fig:arealaw} shows.
Lengthening the cut raises the entropy linearly, well inside the bound of an area law.

\subsection{Range of two-qubit entanglement}
An area law does not say how far apart two entangled qubits can be. However, the two-qubit
reduced states do. Table~\ref{tab:pairs} lists them for every pair on the
$3\times4$ lattice. The insights below can be used to understand Table~\ref{tab:pairs}.
\begin{itemize}
\item \emph{Any set of data qubits is separable.} Tracing out the address register
  leaves $2^{-|\aq|}\sum_x\bigotimes_d\ketbra{\psi^{(d)}_{x_{N(d)}}}{\psi^{(d)}_{x_{N(d)}}}$, a
  mixture of product states. Two data qubits can be classically correlated but are never
  entangled.
\item \emph{A data qubit $d$ and a non-adjacent address qubit $a$ are separable.}
  Tracing out the other address qubits sets their bra and ket configurations equal. Since
  $a\notin N(d)$, the factor on $d$ depends on those labels alone, and $\rho_{da}$ is a
  mixture of products.
\item \emph{Two address qubits with no common data neighbour are separable}, by the same
  argument.
\end{itemize}
Two cases remain. An adjacent data--address pair is generically entangled. A pair of
address qubits sharing a data qubit can be. The second case is easy to miss, because over
random payloads its negativity is tiny. Take one data qubit controlled by two address
qubits, with angle $\pi$ at one of the four address configurations and $0$ at the other
three. Tracing out the data qubit leaves the address pair in
$\tfrac34\ketbra{\omega}{\omega}+\tfrac14\ketbra{e}{e}$, where $\ket{e}$ is the singled-out configuration and $\ket{\omega}$ is the uniform superposition of the other three. This state is
entangled. The data qubit acts as an imperfect which-path record: it cannot tell three of
the four branches apart, so those three stay coherent. Two-qubit entanglement therefore
reaches graph distance two and no further. Classical correlation reaches slightly further
and then vanishes.

\begin{table}[t]
\centering
\caption{Negativity $\mathcal N$ and mutual information $I$ (bits) for every pair on the
$3\times4$ lattice, by pair type (D data, A address) and graph distance $\delta$, over
$40$ random payloads. Entries of $0$ vanish identically.}
\label{tab:pairs}
\footnotesize
\begin{tabular}{lccccc}
\toprule
pair & $\delta$ & mean $\mathcal N$ & max $\mathcal N$ & mean $I$ & max $I$\\
\midrule
DA & 1 & $0.031$     & $0.334$ & $0.189$ & $1.329$\\
AA & 2 & $<10^{-4}$  & $0.009$ & $0.052$ & $0.396$\\
DD & 2 & $0$         & $0$     & $0.018$ & $0.290$\\
DA & 3 & $0$         & $0$     & $0.002$ & $0.070$\\
AA & 4 & $0$         & $0$     & $0.001$ & $0.010$\\
DD & 4 & $0$         & $0$     & $0$     & $0$\\
DA & 5 & $0$         & $0$     & $0$     & $0$\\
\bottomrule
\end{tabular}
\end{table}

\subsection{Effect of shared address qubits}
With a private address register per super-pixel, the state would be a product of
independent small encodings. Shared address qubits join these clusters into one
connected state. For a generic payload every cut that severs a coupler carries some
entanglement, but each shared coupler adds a short bond worth at most one bit. This is
consistent with Proposition~\ref{prop:indep}. Independence of the super-pixels does not mean the state
is a product. It means the entanglement is invisible to the address-diagonal blocks of
$\rho_S$ where the data is encoded. The entanglement lives in the off-diagonal
blocks, i.e. the coherences discussed in App.~\ref{sec:downstream}.

\section{What do the coherences mean?}
\label{sec:downstream}
As a memory, the encoding loses to a classical memory because of the exponential
prefactor of Sec.~\ref{sec:independence}. Full readout is not the only use of a state,
however. If a quantity computed on the encoded state has a small answer, the shot cost
depends on the size of the answer, not of the payload. Proposition~\ref{prop:indep} shows where to
look. The payload is in the address-diagonal blocks of $\rho_S$, and the correlations are
in the off-diagonal blocks. A change of basis on the address register makes the
off-diagonal blocks measurable.

Take the encoded state, apply a Hadamard gate to every address qubit, and measure every qubit in the computational basis. Fix an address qubit $a$ with data
neighbours $\dq_a$, and split the state on its bit,
$\ket{\Psi}=\tfrac{1}{\sqrt2}\big(\ket{0}_a\ket{\Phi_0}+\ket{1}_a\ket{\Phi_1}\big)$. The
Hadamard on $a$ gives $P(z_a=0)=\tfrac12(1+\braket{\Phi_0}{\Phi_1})$; every quantity here
is real. Writing $K_a=2P(z_a=0)-1$ and computing the overlap,
\begin{equation}
  K_a=\mathop{\mathbb{E}}_{x'}\;\prod_{d\in\dq_a}
  \cos\!\left(\frac{\theta^{(d)}_{x',\,x_a=0}-\theta^{(d)}_{x',\,x_a=1}}{2}\right),
  \label{eq:Ka}
\end{equation}
with the expectation uniform over the other $|\aq|-1$ address bits, $x'$. Here $\theta^{(d)}_{x',\,x_a=b}$ is the angle of $d$ at the local address configuration of $N(d)$ in the address string with bit $a$ equal to $b$ and the other bits equal to $x'$. The Hadamards on the
other address qubits do not change this, since a unitary on a subsystem that is then
traced out leaves the marginal unchanged. One gate layer and one basis give every
address qubit's $K_a$ from the same shots.

\subsection{Interpretation}
The two angles in Eq.~\eqref{eq:Ka} are the values stored in super-pixel $d$ at two
addresses that differ only in the bit of $a$. They are neighbouring entries of that
super-pixel along the direction $a$. $K_a$ is an average of products of
cosines of local finite differences. It equals $1$ exactly when flipping the bit $a$
changes nothing, and it falls as the field varies in that direction. It is a smoothness
functional, or edge detector, evaluated with one layer of single-qubit gates. In exact
simulation a constant payload gives $K_a=1$, a smooth ramp stays close to $1$, and a
random binary payload drives $K_a$ towards zero.

\subsection{Shot cost}
$K_a$ comes from a single-qubit marginal, so every shot contributes. Its standard error
is $2\sqrt{P(1-P)/N}\le 1/\sqrt N$, bounded in advance for any degree of $a$. It has no
$2^{m}$ factor. In noiseless sampling the error falls as $1/\sqrt N$. It grows only
slightly with degree, because a higher-degree address qubit has a smaller $K_a$ and so a
$P$ closer to one half, where the binomial variance peaks. 

This is not an exponential saving over the classical alternative, which is to
reconstruct the payload with the ordinary readout and evaluate Eq.~\eqref{eq:Ka} from the
reconstructed angles. Each angle has variance $2^{m}/N$. But $K_a$ averages some
$\Delta_{\aq}2^{m}$ of them, so each enters with weight $O(2^{-m})$, and
$\operatorname{Var}(\hat K_a)\approx\Delta_{\aq}/N$. The $2^{m}$ cancels. Both routes need
$O(1/\epsilon^2)$ shots. Asymptotically the classical route is better, because
conditioning on the address is a stratification, and stratified estimators have lower
variance.

\begin{figure}[h]
  \centering
  \includegraphics[width=\linewidth]{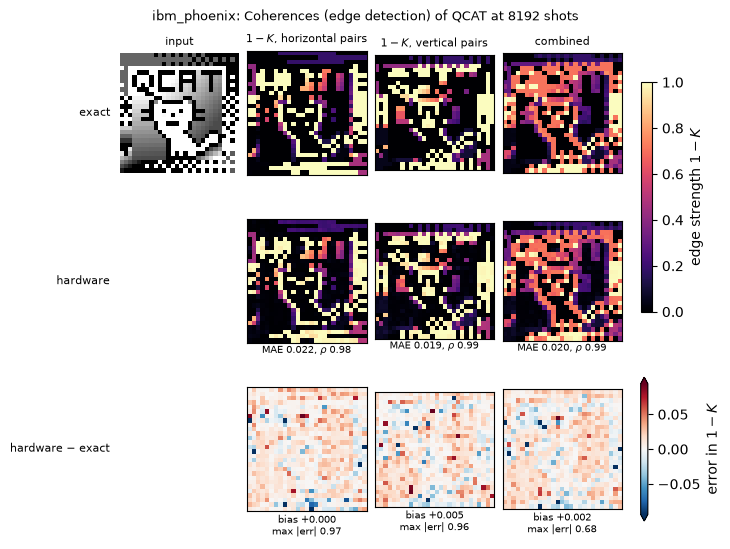}
  \caption{Measured coherences for the large QCAT image on Phoenix at 8192 shots for horizontally and vertically adjacent pixel pairs rather than superpixels for a comprehensible visualization of the procedure.}\label{fig:coherences}
\end{figure}

The interference route has two advantages. The classical route inverts a nonlinear
function of each per-configuration frequency, which gives a bias of order $2^{m}/N$. It
is also undefined when an address configuration receives no shots, which is common for
$N\lesssim 2^{m}$. The interference route is unbiased and always defined.
Table~\ref{tab:kcompare} shows the crossover and Fig.~\ref{fig:coherences} shows the results on the QCAT image on Phoenix distributed so that the coherences are measured between pixel pairs instead of superpixel pairs. At a few dozen shots the classical route
has a bias larger than its standard error and fails in a quarter of repetitions. Above a
few thousand shots it is more accurate. The interference route is preferable when shots
are scarce relative to $2^{m}$, which would be typical on higher-degree devices; when a
guaranteed error bar is needed; and when the payload itself is not needed.

\begin{table}[t]
\centering
\caption{Estimating $K_a$ for a degree-four address qubit on the $3\times4$ lattice
($200$ repetitions, noiseless sampling): the interference route of Eq.~\eqref{eq:Ka}
against reconstructing the payload and evaluating Eq.~\eqref{eq:Ka} classically.
``Undefined'' is the fraction of repetitions in which some address configuration
received no shots.}
\label{tab:kcompare}
\footnotesize\setlength{\tabcolsep}{4pt}
\begin{tabular}{lcccc}
\toprule
 & interference & \multicolumn{3}{c}{classical reconstruction}\\
\cmidrule(lr){2-2}\cmidrule(lr){3-5}
shots $N$ & RMS error & RMS error & bias & undefined\\
\midrule
$64$      & \textbf{0.113} & 0.169 & $-0.158$ & $26\%$\\
$256$     & \textbf{0.054} & 0.056 & $-0.044$ & $0\%$\\
$1\,024$  & 0.029 & \textbf{0.022} & $-0.015$ & $0\%$\\
$4\,096$  & 0.015 & \textbf{0.010} & $-0.006$ & $0\%$\\
$65\,536$ & 0.004 & \textbf{0.003} & $-0.001$ & $0\%$\\
\bottomrule
\end{tabular}
\end{table}

\subsection{Limitations}
The coherence readout is a separate job. The Hadamards destroy the address-diagonal
information, so the same shots cannot give both the payload and its coherences. There is
no asymptotic advantage: a classical computer can evaluate $K_a$ from the payload in time
linear in the capacity. It can be estimated from the ordinary readout at comparable or lower
shot cost. Table~\ref{tab:kcompare} comes from simulation; Fig.~\ref{fig:coherences} shows a related hardware run. A
constant payload gives $K_a=1$ exactly, so on hardware its deviation from $1$ would
measure device decoherence directly, with no reference measurement.

The functional itself is modest. The broader point is that the encoded state supports
operations with small outputs that cost one layer of single-qubit gates. Better
operations of this kind may exist.

\section{Full single qubit tomography for more precision or one more bit of data}
\label{sec:xyz}
The $X$ basis measurement gives $\cos\varphi$ and cannot distinguish $\varphi$ from
$2\pi-\varphi$. This limits the second field to a half circle. A $Y$ measurement supplies
$\sin\varphi$ and removes the ambiguity. There are two ways to use it. Throughout this appendix, $n\approx N/2^{m_d}$ is the number of shots per local address configuration in each basis, and $c$ is the polar clamp of App.~\ref{sec:clamp}: we map the first value to $\theta\in[c,\pi-c]$.

The first is to extend the second field to the full circle, $\varphi\in[0,2\pi)$,
decoded by
\begin{equation}
  \begin{aligned}
  \theta_i&=\operatorname{atan2}\!\big(\textstyle\sqrt{\langle X\rangle_i^2+\langle Y\rangle_i^2},\,\langle Z\rangle_i\big),\\
  \varphi_i&=\operatorname{atan2}\!\big(\langle Y\rangle_i,\langle X\rangle_i\big).
  \end{aligned}
  \label{eq:decodefull}
\end{equation}
This doubles the range, and the $\operatorname{atan2}$ has no steep ends:
\begin{equation}
  \sigma_\varphi^2=\frac{1-2\sin^2\!\theta\,\cos^2\!\varphi\,\sin^2\!\varphi}{n\,\sin^2\!\theta}
  \le\frac{1}{n\sin^2\!\theta}.
  \label{eq:sigmafull}
\end{equation}
Only the polar clamp remains in the denominator. At the default clamp the second field's
error is about three times lower than on the half circle, and lower than that of the
single encoding. The first field also improves slightly, since all three components now
inform it (Fig.~\ref{fig:pairclamp}). The full circle has one drawback: $\varphi=0$ and
$2\pi$ are the same point. Noise that moves a value across this point scrambles a
non-periodic field whose extremes meet there. A clamp $c_b$ on $\varphi$ leaves a gap
around the join and prevents this. For a periodic field, such as a hue, a phase or a
direction, the join is correct and $c_b=0$.

The second way keeps the half circle and stores an extra bit in the sign of $\varphi$.
Encode a continuous $\varphi_0\in[c_b,\pi-c_b]$ and a bit $s$ as $\varphi=\varphi_0$ for
$s=0$ and $\varphi=2\pi-\varphi_0$ for $s=1$. The $X$ basis does not see $s$, since the
cosine is even, and returns $\varphi_0$ as before. The $Y$ basis returns $s$ as the sign
of $\langle Y\rangle$. Each slot then carries two continuous values and one bit. Both
clamps are required: the sign of $\sin\varphi$ is undefined at $\varphi=0$ and $\pi$, and
vanishes with $\sin\theta$ at the poles. With both clamps,
$|\langle Y\rangle|\ge\sin c\,\sin c_b$. The estimate of $\langle Y\rangle$ has standard
deviation at most $n^{-1/2}$, so the probability of a wrong bit is at most
$\tfrac12\operatorname{erfc}\!\big(\sqrt{n/2}\,\sin c\,\sin c_b\big)$, the normal tail probability.
Without noise this is negligible at any practical shot count. The bound also gives the
clamp needed for a given noise level.

The choice depends on the data. Periodic fields, and fields that need resolution, suit
the full circle. Payloads with a natural binary label, such as a mask, a flag or a sign,
suit the extra bit. If neither is needed, the $Y$ basis can be dropped.
Table~\ref{tab:modes} compares the options.

\begin{table}[t]
\centering
\caption{The three readouts. Shots per value are relative to the single encoding at a
fixed number of shots per basis. The last two rows are one readout, used either for a
full circle or for a half circle and a bit per slot.}
\label{tab:modes}
\footnotesize\setlength{\tabcolsep}{3pt}
\begin{tabular}{lcccc}
\toprule
Bases & Values & $\varphi$ & Clamped & Shots/Value\\
\midrule
$Z$       & $\mathcal C$  & ---              & ---               & $1$\\
$Z,X$     & $2\mathcal C$ & $[0,\pi]$        & $\theta$          & $1$\\
$Z,X,Y$   & $2\mathcal C$ & $[0,2\pi)$       & $\theta$, $\varphi^{*}$ & $1.5$\\
$Z,X,Y$   & $2\mathcal C$\,+\,$\mathcal C$ bits & $[0,\pi]$\,+\,sign & $\theta$, $\varphi$ & $1.5$\\
\bottomrule
\multicolumn{5}{l}{\scriptsize $^{*}$only if the field is not periodic.}
\end{tabular}
\end{table}

\subsection{Clamping the polar angle}
\label{sec:clamp}
\begin{figure}[t]
\centering
\includegraphics{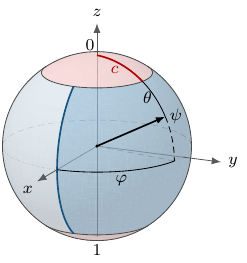}
\caption{The clamped encoding region on one qubit, for the default $c=0.2\pi$. The first
value sets the polar angle, restricted to $\theta\in[c,\pi-c]$. The pink caps near the
poles are excluded, because the transverse component that carries $\varphi$ cannot be read out reliably. The blue patch, $\varphi\in[0,\pi]$, is the region the $Z,X$ readout can
address; its edge at $\varphi=0$ is drawn dark. The pale band $\varphi\in[\pi,2\pi)$ is
added by the $Y$ basis of Sec.~\ref{sec:xyz}. The arrow is an example state.}
\label{fig:clampregion}
\end{figure}

A slot at $\theta=0$ or $\pi$ is $\ket0$ or $\ket1$ and has no azimuth, so a second value
written there is lost. Near the poles the signal is weak, because $\langle X\rangle$ and
$\langle Y\rangle$ scale with $\sin\theta$. We therefore map the first set of numbers onto
$[c,\pi-c]$ instead of $[0,\pi]$. This shrinks the first set of numbers' range from $\pi$ to
$\pi-2c$. Figure~\ref{fig:clampregion} shows the remaining region of the sphere.

\begin{figure*}[h]
\centering
\includegraphics[width=0.82\linewidth]{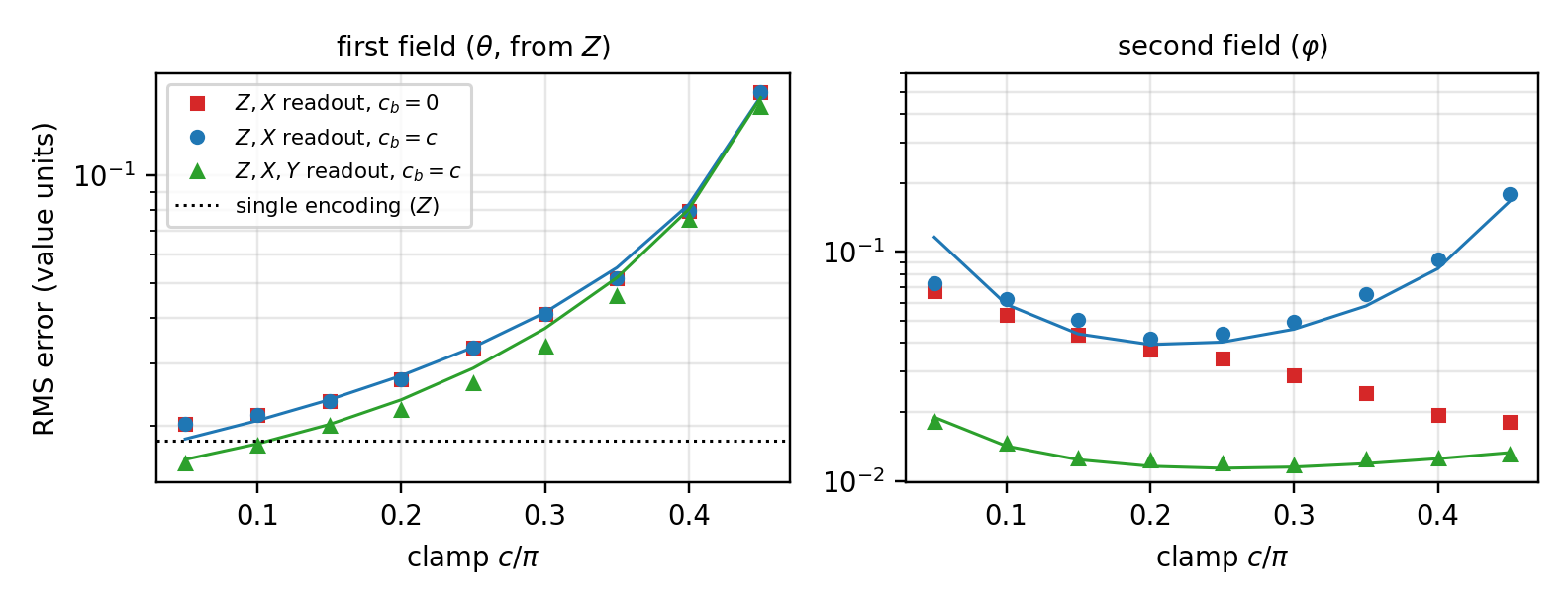}
\caption{The clamp tradeoff under shot noise only: $4096$ shots per basis on a
$4\times4$ lattice, three random payloads per point. Markers are sampled RMS errors in
data units; lines are Eqs.~\eqref{eq:sigmapair} and~\eqref{eq:sigmafull}. Left: the first
set of numbers loses resolution as its range $\pi-2c$ shrinks; the dotted line is the unclamped
single encoding. Right: the second set of numbers. On the half circle ($Z,X$) it is best near
$c\approx0.2\pi$ when $\varphi$ is also clamped ($c_b=c$), and keeps improving with $c$
when it is not ($c_b=0$)}
\label{fig:pairclamp}
\end{figure*}

To understand the error tradeoff, let each address configuration of a degree-$m$
slot receive $n\approx N/2^m$ of the $N$ shots per basis. Then
\begin{equation}
  \sigma_\theta^2=\frac1n,\qquad
  \sigma_\varphi^2=\frac{1+\cos^2\!\varphi\,\cos2\theta}{n\,\sin^2\!\theta\,\sin^2\!\varphi}.
  \label{eq:sigmapair}
\end{equation}
The first term is the $2^{m/2}/\sqrt N$ of Sec.~\ref{sec:independence} and does not depend on
$\theta$. The second contains the factor $1/\sin^2\theta$ which is clamped. In
data units the first set of numbers' error, $\sigma_\theta/(\pi-2c)$, rises with $c$, and the
second set of numbers' error falls. Figure~\ref{fig:pairclamp} compares sampled errors with
Eq.~\eqref{eq:sigmapair}. They agree except at the smallest clamp, where the linear
approximation fails.

Clamping $\varphi$ does not help. The factor $1/\sin^2\varphi$ in
Eq.~\eqref{eq:sigmapair} diverges at the ends of the half circle, where the $\arccos$ is
steepest. The divergence is an artefact of linearising a bounded function. The estimate
stays in $[0,\pi]$ and converges more slowly there, as $n^{-1/4}$, but does not fail. In
sampling, leaving $\varphi$ unclamped gives lower error since clamping gives up range
for an unnecessary margin.
We use $c=0.2\pi$, which seems to be the optimal tradeoff. The start of this appendix describes the extra bit and the full single-qubit tomography. The second channel is not free: we must also measure the state in a second basis, although the mathematics to recover the values is the same. For the same precision per value, the total shot count doubles, but the shot count per value stays the same.

\section{Practical considerations}
\label{sec:considerations}

\subsection{Transforming the data}
Tessa (and its related encodings) spreads its resolution evenly across the angle range. It reconstructs a field
best when the values are spread evenly over their range. Heavy-tailed fields, such as
note durations, where most notes are short and a few are long, crowd most values into a
small interval. There, noise of fixed scale erases the detail. An invertible, monotonic
transform $w$ is applied before normalisation and inverted after decoding. 
The best transform for a dataset must be found by trial or knowledge of the data structure, 
although in a finite alphabet the mapping can be made linear trivially by placing the symbols at equidistant points.
Of course, in a discrete representation one loses the continuous transition between points.

\subsection{Dynamic-range recovery}
\label{sec:drr}
Noise compresses the reconstruction towards its centre and reduces contrast. Rescaling
towards the spread of the input recovers much of it. The rescaling may use a few summary
statistics of the input, never the input itself. Table~\ref{tab:drr} lists the methods
and their effect on the hardware images of Sec.~\ref{sec:hwrun} and the MIDI of
App.~\ref{app:midi}.
QCrank used similar classical post-processing~\cite{qcrank}. Here we compare several methods.

Matching the mean and standard deviation (\texttt{mean\_std}) of the input worked best.
 Matching the minimum and maximum (\texttt{min\_max}) depends
on the extreme values but can be done blind since the data is normalised. 
Under noise, however, that value can be an outlier, and the method then
stretches the whole reconstruction to fit an error. A percentile variant gives up a little
range in exchange for resistance to outliers. The \texttt{min\_avg\_max} method combines
an affine stretch fixing the minimum and maximum with a monotone gamma,
$a'\mapsto a_{\min}+(a_{\max}-a_{\min})\,u^{\gamma}$ for $u\in[0,1]$, with $\gamma$ chosen
so that the mean also matches. It pins the same endpoints as \texttt{min\_max} but does
much better on MIDI. A single multiplier
(\texttt{abs\_max}) works well for signals anchored at zero that still return a 0 value after decoding, but then it is equivalent to \texttt{min\_max}.

The choice of method matters, and different metrics can disagree. On MIDI,
\texttt{min\_max} lowers the mean error but recovers fewer exact pitches and durations than no
correction. Stretching moves most values slightly closer to their targets, which lowers
the mean error. It also pushes many values across a semitone boundary, which lowers the
exact count. On the images, fixing the endpoints also did worse than matching the \texttt{mean\_std}.
The best method may depend on the payload, but in every test here the \texttt{mean\_std} was a better
guide than the extremes. 

\begin{table}[t]
\centering
\caption{Dynamic-range recovery on hardware. ``Scalars'' counts the summary numbers of
the input each method uses. Image columns give the mean absolute error of the $Z$ runs of
Sec.~\ref{sec:hwrun}. Pitch columns give the exact-semitone rate and mean error of the
Fez score of App.~\ref{app:midi}.}
\label{tab:drr}
\footnotesize\setlength{\tabcolsep}{2.5pt}
\begin{tabular}{llcccc}
\toprule
 & & \multicolumn{2}{c}{image MAE} & \multicolumn{2}{c}{pitch}\\
\cmidrule(lr){3-4}\cmidrule(lr){5-6}
method & scalars & Phoenix & Fez & exact (\%) & MAE (st)\\
\midrule
none                  & 0 & 0.130 & 0.131 & 18.7 & 2.13\\
\texttt{abs\_max}     & 1 & 0.120 & 0.121 & ---  & ---\\
\texttt{min\_max}     & 2 & 0.114 & 0.086 & 8.4  & 1.83\\
\texttt{percentile}   & 2 & 0.085 & \textbf{0.076} & 12.4 & 1.67\\
\texttt{mean\_std}    & 2 & \textbf{0.075} & 0.078 & \textbf{49.3} & \textbf{0.66}\\
\texttt{min\_avg\_max}& 3 & 0.112 & 0.090 & 23.3 & 1.13\\
\bottomrule
\end{tabular}
\end{table}

\subsection{Quantisation onto a known alphabet}
\label{app:quantise}
If the data lie on a known discrete set, such as a palette, a set of permitted durations
or a codebook, rounding the reconstruction onto that set is free. It uses no information
about the particular values. Rounding makes the error discrete. Errors below half the
spacing disappear; larger errors become wrong values. The reconstruction error reduces to
asking ``what fraction of values moved by more than half a bin?'' This explains the
music and genetic results of Appendices \ref{app:midi} and \ref{app:genetic}, where decoded symbols are either exactly right or
clearly wrong. The method works only while the spacing exceeds the reconstruction error.
On current hardware this limits it to alphabets of a few levels. However, for things like indexed colour, where each value is attached to a palette value, it is possible to encode and decode coloured pixel art.

\section{Other applications}
\label{app:music}
The encoding does not depend on what the numbers represent. This appendix collects various modes of operation for pixel art, sound and MIDI.

\subsection{Pixel art}

Pixel art can be encoded on a quantum computer via Tessa in a variety of ways. One can use grayscale values directly, allowing for the largest images, of course. However, for colour there are multiple approaches. One can use a pair representation of colour, such as hue and value as is done in the main text and Fig.~\ref{fig:pixelart}(a) on \emph{City-pop Cat}, where the hue and value are encoded into one slot of the Tessa encoding, which can hold a value in the Z basis and another in the (X,Y) direction of each reduced density matrix that can be read out. If one were to desire full colour (RGB, HSV, etc.) then one can distribute the data over more `slots' as has been done in \emph{Portrait} (RGB) and \emph{Ramen} (HSV). Alternatively one can also use indexed colour where one partitions the values into areas that are clipped to specific indices in a palette as is done in \emph{Spaceship}. 

Another interesting thing that can be done is encoding a pictorial cat state as in Fig.~\ref{fig:pixelart}(b), where the encoding `slots' have the image of a living cat in the $Z$ basis and of a dead cat in the $X$ basis. If one then measures at different angles one will recover a mixture of the two, and measuring $Z$ or ($X$, $Y$) directly ought to, without the presence of noise, recover the completely alive and completely dead cat. This is shown in \emph{QCAT} and for a less depressing example, with a moth and apple. At least from the moth's point of view. This can also be done with indexed colour if the palette is shared between the two as in the lowest example in the figure.

\subsection{Wavetable synthesis}
\label{app:wavetable}
\begin{figure}[h]
\centering
\includegraphics[width=\linewidth]{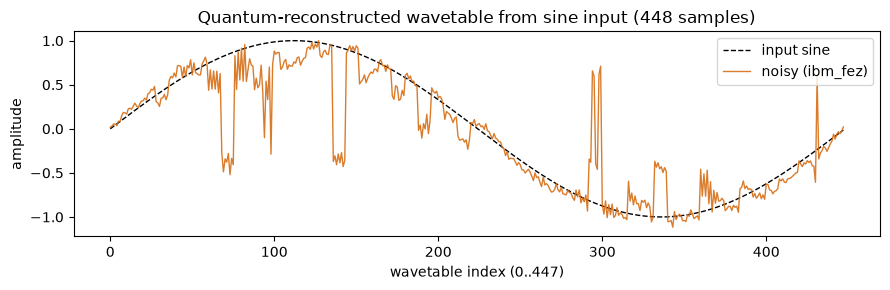}
\caption{A $448$-sample single-cycle sine (dashed) and the waveform recovered from
Fez at $8192$ shots after \texttt{min\_max} dynamic-range recovery (solid). The noise-driven
excursions change the synthesised timbre.}
\label{fig:sound}
\end{figure}

A single waveform cycle stored as the payload can serve as a wavetable. An oscillator
scans the reconstructed table at each note's frequency, so the waveform sets the timbre
and the scan rate sets the pitch. We encoded a $448$-sample sine, read it back, and
rendered a MIDI piece (Bach's Invention~15) with it, which can be found in the code. The noiseless reconstruction gives a
nearly pure tone. The hardware reconstruction (Fig.~\ref{fig:sound}), after
dynamic-range recovery, follows the sine with small excursions. These add overtone energy
and audibly change the timbre. This is a defect, but in this application the output
remains musically usable.

\subsection{Symbolic music (MIDI)}
\label{app:midi}
One may, for example, encode a score as two lists, per-note pitch and duration, each in its own run. After
decoding one ought to round the pitches and durations to the nearest semitone or dotted stem and rebuild the score with the
original onsets. The test piece is an excerpt from Telemann's twelfth fantasia for solo
flute: $347$ notes against a capacity of $448$. Without noise all MIDI data is
recovered exactly.

Rounding makes pitch unforgiving. An error below half a semitone vanishes; an error of a
whole semitone is wrong. The raw hardware run recovers fewer than half of the pitches
exactly. The large gain from two summary statistics and one transformation function 
is the more useful result. This is shown in simulation and hardware in Table~\ref{tab:midi}.

At this fidelity the output can serve as a set of variations on a theme. The melodic
contour survives while individual pitches and durations change, and the device's noise
supplies the variation. The scheme is not a lossless channel for symbolic music. A more
serious application of the same method is encoding DNA or protein sequences.
Quantisation onto an alphabet and the encoding of biological sequences have both been
studied before~\cite{qcrank} and follows this appendix.

\begin{table}[h]
\centering
\caption{Symbolic music on Fez. ``Exact'' is the percentage of pitches
recovered to the correct semitone; pitch and duration errors are mean absolute errors.
Recovered runs apply \texttt{mean\_std} to the pitches and the named transformation to the
durations.}
\label{tab:midi}
\footnotesize\setlength{\tabcolsep}{1pt}
\begin{tabular}{lccc}
\toprule
 & exact (\%) & pitch (st) & dur.\ (s)\\
\midrule
noiseless simulation                      & 100.0 & 0.00 & 0.000\\
noisy simulation, raw                     & 18.7  & 2.13 & 0.491\\
noisy simulation, linear                  & 49.3  & 0.66 & 0.208\\
noisy simulation, $\log$                  & 49.3  & 0.66 & 0.145\\
noisy simulation, square root             & 49.3  & 0.66 & 0.117\\
hardware (Fez), \texttt{min\_max}, linear & 39.5  & 1.62 & 0.469\\
\bottomrule
\end{tabular}
\end{table}

\makeatletter
\makeatother

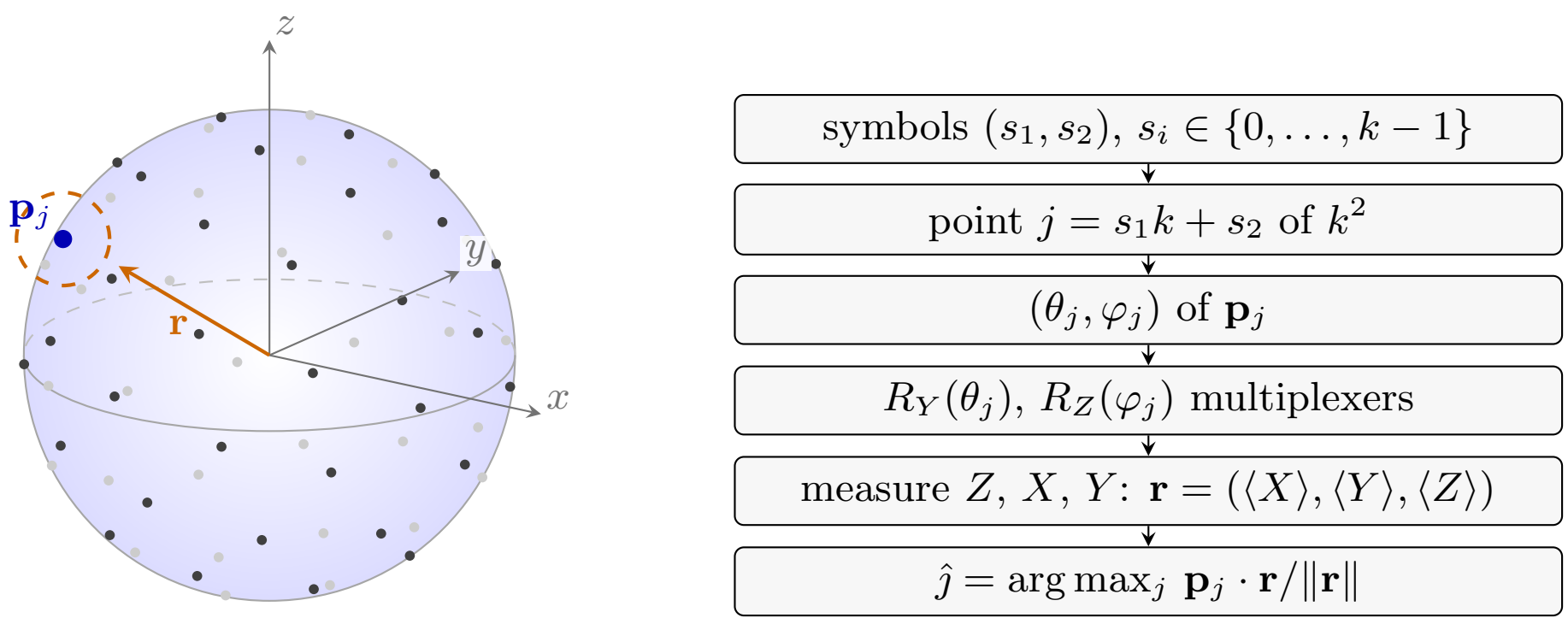
\begin{figure*}[t]
  \centering
\begin{tikzpicture}[>=stealth, font=\small]
\begin{scope}
\shade[inner color=white, outer color=blue!14] (0,0) circle (1.9);
\draw[gray!70] (0,0) circle (1.9);
\draw[gray!50, dashed] (1.9,0) arc (0:180:1.9 and 0.587);
\draw[gray!70] (-1.9,0) arc (180:360:1.9 and 0.587);
\fill[fill=black!75] (-0.076,1.586) circle (1.1pt);
\fill[fill=black!75] (-0.373,1.841) circle (1.1pt);
\fill[fill=black!75] (0.616,1.709) circle (1.1pt);
\fill[fill=black!75] (-0.992,1.384) circle (1.1pt);
\fill[fill=black!20] (0.317,1.858) circle (1.1pt);
\fill[fill=black!75] (0.627,1.256) circle (1.1pt);
\fill[fill=black!75] (-1.177,1.491) circle (1.1pt);
\fill[fill=black!75] (1.279,1.402) circle (1.1pt);
\fill[fill=black!75] (-0.505,1.012) circle (1.1pt);
\fill[fill=black!20] (-0.469,1.758) circle (1.1pt);
\fill[fill=black!75] (1.338,1.030) circle (1.1pt);
\fill[fill=black!20] (0.953,1.486) circle (1.1pt);
\fill[fill=black!75] (0.173,0.697) circle (1.1pt);
\fill[fill=black!20] (-1.229,1.219) circle (1.1pt);
\fill[fill=black!75] (1.752,0.706) circle (1.1pt);
\fill[fill=black!75] (-1.220,0.592) circle (1.1pt);
\fill[fill=black!20] (0.247,1.506) circle (1.1pt);
\fill[fill=black!75] (1.027,0.425) circle (1.1pt);
\fill[fill=black!20] (-1.735,0.700) circle (1.1pt);
\fill[fill=black!20] (1.527,0.891) circle (1.1pt);
\fill[fill=black!75] (-0.545,0.164) circle (1.1pt);
\fill[fill=black!20] (-0.548,1.256) circle (1.1pt);
\fill[fill=black!75] (1.613,0.174) circle (1.1pt);
\fill[fill=black!75] (-1.694,0.110) circle (1.1pt);
\fill[fill=black!20] (0.914,0.929) circle (1.1pt);
\fill[fill=black!75] (0.336,-0.138) circle (1.1pt);
\fill[fill=black!20] (-1.455,0.510) circle (1.1pt);
\fill[fill=black!20] (1.830,0.115) circle (1.1pt);
\fill[fill=black!75] (-1.198,-0.319) circle (1.1pt);
\fill[fill=black!20] (0.097,0.795) circle (1.1pt);
\fill[fill=black!75] (1.169,-0.408) circle (1.1pt);
\fill[fill=black!75] (-1.898,-0.070) circle (1.1pt);
\fill[fill=black!20] (1.392,0.181) circle (1.1pt);
\fill[fill=black!75] (-0.371,-0.708) circle (1.1pt);
\fill[fill=black!20] (-0.772,0.578) circle (1.1pt);
\fill[fill=black!75] (1.862,-0.244) circle (1.1pt);
\fill[fill=black!75] (-1.615,-0.700) circle (1.1pt);
\fill[fill=black!20] (0.655,0.099) circle (1.1pt);
\fill[fill=black!75] (0.478,-0.907) circle (1.1pt);
\fill[fill=black!20] (-1.710,-0.237) circle (1.1pt);
\fill[fill=black!20] (1.614,-0.558) circle (1.1pt);
\fill[fill=black!75] (-0.945,-1.140) circle (1.1pt);
\fill[fill=black!20] (-0.249,-0.053) circle (1.1pt);
\fill[fill=black!75] (1.512,-0.846) circle (1.1pt);
\fill[fill=black!20] (-1.683,-0.854) circle (1.1pt);
\fill[fill=black!20] (1.033,-0.667) circle (1.1pt);
\fill[fill=black!75] (-0.059,-1.430) circle (1.1pt);
\fill[fill=black!20] (-1.098,-0.277) circle (1.1pt);
\fill[fill=black!20] (1.647,-0.945) circle (1.1pt);
\fill[fill=black!75] (-1.235,-1.392) circle (1.1pt);
\fill[fill=black!20] (0.264,-0.688) circle (1.1pt);
\fill[fill=black!75] (0.864,-1.380) circle (1.1pt);
\fill[fill=black!20] (-1.244,-0.969) circle (1.1pt);
\fill[fill=black!20] (1.120,-1.331) circle (1.1pt);
\fill[fill=black!75] (-0.559,-1.708) circle (1.1pt);
\fill[fill=black!20] (-0.549,-0.924) circle (1.1pt);
\fill[fill=black!75] (1.086,-1.551) circle (1.1pt);
\fill[fill=black!20] (-1.039,-1.526) circle (1.1pt);
\fill[fill=black!20] (0.418,-1.377) circle (1.1pt);
\fill[fill=black!75] (0.343,-1.810) circle (1.1pt);
\fill[fill=black!20] (-0.393,-1.563) circle (1.1pt);
\fill[fill=black!20] (0.468,-1.779) circle (1.1pt);
\fill[fill=black!20] (-0.339,-1.858) circle (1.1pt);
\draw[->, black!55] (0,0) -- (0.000,2.439) node[anchor=south west, inner sep=1pt, fill=white, fill opacity=0.8, text opacity=1] {$z$};
\draw[->, black!55] (0,0) -- (2.101,-0.455) node[anchor=south west, inner sep=1pt, fill=white, fill opacity=0.8, text opacity=1] {$x$};
\draw[->, black!55] (0,0) -- (1.471,0.649) node[anchor=south west, inner sep=1pt, fill=white, fill opacity=0.8, text opacity=1] {$y$};
\draw[orange!80!black, thick, dashed] (-1.597,0.899) circle (0.36);
\draw[->, thick, orange!80!black] (0,0) -- (-1.167,0.690);
\node[orange!80!black, anchor=north east, inner sep=1pt] at (-0.584,0.345) {$\mathbf{r}$};
\fill[blue!70!black] (-1.597,0.899) circle (2pt);
\node[blue!70!black, anchor=south east, inner sep=2pt] at (-1.597,0.899) {$\mathbf{p}_j$};
\end{scope}
\node[draw, rounded corners=2pt, fill=gray!6, minimum width=6.4cm, inner sep=3pt] (s0) at (6.800000000000001,1.75) {symbols $(s_1, s_2)$, $s_i \in \{0,\dots,k-1\}$};
\node[draw, rounded corners=2pt, fill=gray!6, minimum width=6.4cm, inner sep=3pt] (s1) at (6.800000000000001,1.05) {point $j = s_1 k + s_2$ of $k^2$};
\draw[->] (s0) -- (s1);
\node[draw, rounded corners=2pt, fill=gray!6, minimum width=6.4cm, inner sep=3pt] (s2) at (6.800000000000001,0.35) {$(\theta_j, \varphi_j)$ of $\mathbf{p}_j$};
\draw[->] (s1) -- (s2);
\node[draw, rounded corners=2pt, fill=gray!6, minimum width=6.4cm, inner sep=3pt] (s3) at (6.800000000000001,-0.35) {$R_Y(\theta_j)$, $R_Z(\varphi_j)$ multiplexers};
\draw[->] (s2) -- (s3);
\node[draw, rounded corners=2pt, fill=gray!6, minimum width=6.4cm, inner sep=3pt] (s4) at (6.800000000000001,-1.05) {measure $Z$, $X$, $Y$: $\mathbf{r} = (\langle X\rangle, \langle Y\rangle, \langle Z\rangle)$};
\draw[->] (s3) -- (s4);
\node[draw, rounded corners=2pt, fill=gray!6, minimum width=6.4cm, inner sep=3pt] (s5) at (6.800000000000001,-1.75) {$\hat{\jmath} = \arg\max_j \, \mathbf{p}_j \cdot \mathbf{r} / \lVert \mathbf{r} \rVert$};
\draw[->] (s4) -- (s5);
\end{tikzpicture}
  \caption{The constellation. The symbols of a slot, here two, $(s_1, s_2)$, select
  one of $k^2$ points $\mathbf{p}_j$, spread evenly over the Bloch sphere. The slot is
  prepared at $\mathbf{p}_j$ by the same $R_Y$ and $R_Z$ multiplexers as the pair mode,
  and read in $Z$, $X$ and $Y$. This is an idea derived from \cite{P_rez_Salinas_2020}. The decoder normalises the measured Bloch vector
  $\mathbf{r}$, which noise shrinks, and returns the nearest point (dashed: its
  decision region). Shown: $k = 8$, 64 points. Two amino acids per slot make 400
  points, one makes 20.}
  \label{fig:constellation}
\end{figure*}
\subsection{Encoding DNA and protein sequences}
\label{app:genetic}

To further test the potential applications of Tessa, we stored and retrieved DNA and protein sequences on
Phoenix. These sequences were chosen because their lengths come
closest to the maximum capacities or because of their names, not for their biology. At two symbols per slot they are
the RefSeq mRNA of \emph{POLG2} (NM\_007215.4, 1582 nucleotides)~\citep{OLeary2016}
and isoform~2 of the sulfonylurea receptor SUR1, encoded by \emph{ABCC8} (UniProt
Q09428-2, amino acid sequence of length 1582)~\citep{UniProt2023}. \emph{POLG2} encodes p55, the accessory
subunit of mitochondrial DNA polymerase~$\gamma$~\citep{Lim1999}, whose mutations
cause progressive external ophthalmoplegia~\citep{Longley2006,Young2016}. SUR1 is the
regulatory subunit of the ATP-sensitive potassium channel of pancreatic $\beta$
cells~\citep{AguilarBryan1995}, and mutations in \emph{ABCC8} cause congenital
hyperinsulinism~\citep{Thomas1995} or neonatal diabetes~\citep{Babenko2006}. At one
symbol per slot we chose to use \emph{MT-CO3}, the gene of subunit III of cytochrome
$c$ oxidase in the human mitochondrial genome (NC\_012920.1, positions 9207--9990, 784
nucleotides)~\citep{Anderson1981,Andrews1999}, and RRM1, the large subunit of
ribonucleotide reductase, which makes the deoxyribonucleotides for DNA synthesis
(P23921, amino acid sequence of length 792)~\citep{Nordlund2006}. 

Three shorter human proteins fill part of
the chip: the tumour suppressor p53 (P04637, amino acid sequence of length 393), the ``guardian of the
genome''~\citep{Lane1992,Levine1997}; pikachurin (Q63HQ2, amino acid sequence of length 1017), a
dystroglycan ligand needed for photoreceptor ribbon synapses~\citep{Sato2008}; and
Sonic hedgehog (Q15465, amino acid sequence of length 462), a morphogen that patterns the limb and the
neural tube~\citep{Riddle1993,Echelard1993}. The last two are named after the
video-game characters Pikachu and Sonic the Hedgehog. Using shorter sequences allows one to choose the best qubits on the chip or would allow for other qubits to be used for other computations on the encoded data.

Each symbol is an integer, from an alphabet of $k = 4$ bases or $k = 20$ amino acids.
A sequence of at most 792 symbols takes one symbol per slot. A longer one takes two,
the first half of the sequence in the first position of every slot and the second half
in the second. We compare two encodings. In the \emph{angle} encoding, as for images in
QPIXL~\citep{qpixl} and QCrank~\citep{qcrank}, each symbol is one of $k$
evenly spaced angles: with one symbol per slot, the amplitude $\theta$ over
$[0, \pi]$, read in $Z$ only; with two, $\theta$ and the phase $\varphi$, each kept
$0.2\pi$ from the ends of its range and read in $Z$ and $X$. In the
\emph{constellation} (Fig.~\ref{fig:constellation}), the symbols of a slot are one
point of a constellation of $k$ states, or $k^2$ for a pair $(s_1, s_2)$, which is point
$j = s_1 k + s_2$. The points are a Fibonacci lattice on the whole Bloch
sphere~\citep{Gonzalez2010,SaffKuijlaars1997}, spread further apart by a short mutual
repulsion. Point $j$ sets the angles $(\theta_j, \varphi_j)$, so the circuit is the
full-circle pair mode with the same two-qubit gates. The decoder measures $Z$, $X$ and
$Y$, normalises $\mathbf{r} = (\langle X\rangle, \langle Y\rangle, \langle Z\rangle)$
and returns $\hat{\jmath} = \arg\max_j \mathbf{p}_j \cdot \mathbf{r}/\lVert\mathbf{r}\rVert$.
What determines the error is how far apart the allowed states are. The 400 points of two amino acids are at least $0.16$~rad apart, and 20 angles on the clamped amplitude are
$0.099$~rad apart. Normalising $\mathbf{r}$ also removes any uniform shrinking from
depolarising noise. Dynamic range recovery acts on each angle, or on each
component of $\mathbf{r}$.

\begin{figure*}[t]
  \centering
  \includegraphics[width=\linewidth*8/10]{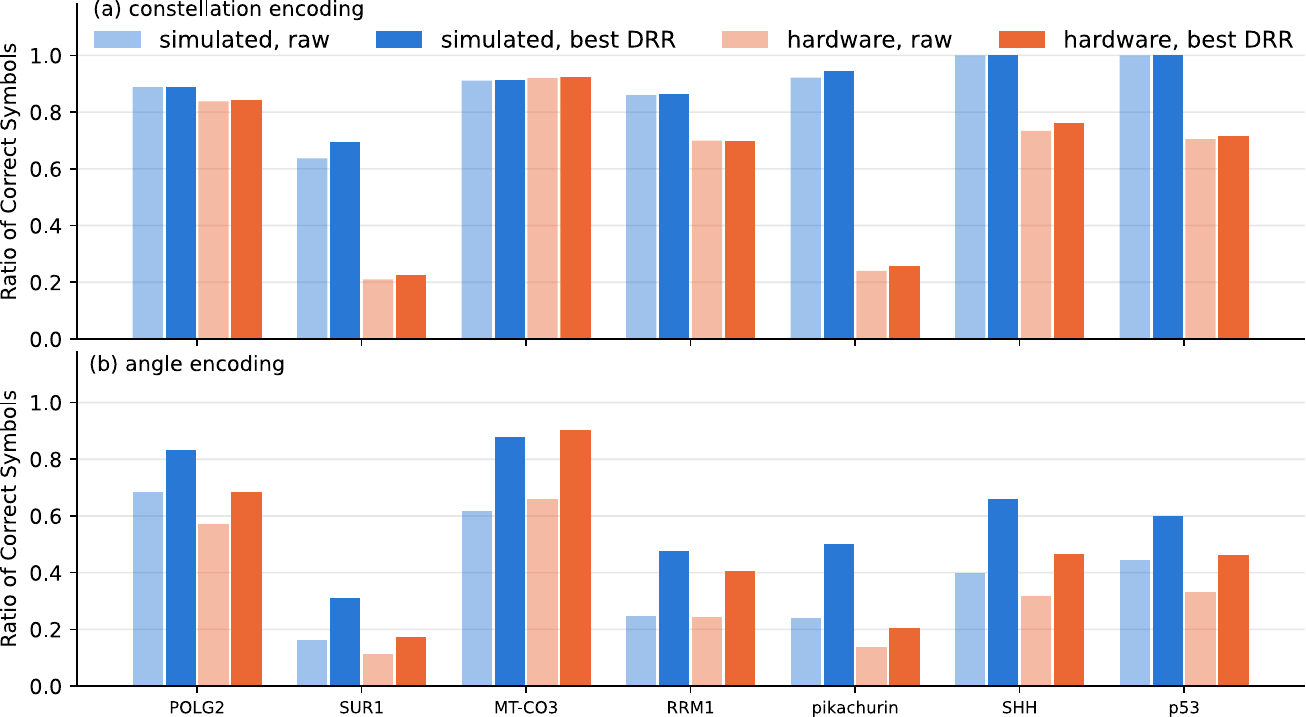}
  \caption{Ratio of correctly decoded symbols to sequence length from a simulated and hardware IBM Phoenix quantum computer
  at 32,768 shots per experiment, raw and with the best dynamic range recovery, for (a) the
  constellation and (b) the angle encoding. SHH is short for Sonic Hedgehog, other proteins are defined in the text.}
  \label{fig:genetic-results}
\end{figure*}

\begin{table*}[t]
  \centering
  \footnotesize
  \setlength{\tabcolsep}{2.2pt}
  \begin{tabular}{l l c c ccc ccc}
    \toprule
    & & & no noise & \multicolumn{3}{c}{simulated noise} & \multicolumn{3}{c}{hardware} \\
    \cmidrule(lr){4-4} \cmidrule(lr){5-7} \cmidrule(lr){8-10}
    sequence & encoding & per slot & raw & raw & min--max & best DRR & raw & min--max & best DRR \\
    \midrule
    POLG2 mRNA & constellation & 2 & 100.0 & 88.8 & 88.3 & 89.0 (ms) & 83.8 & 84.3 & 84.3 (mm) \\
SUR1 isoform 2 & constellation & 2 & 100.0 & 63.7 & 63.0 & 69.5 (ms) & 21.0 & 21.5 & 22.6 (pc) \\
MT-CO3 & constellation & 1 & 100.0 & 91.1 & 91.5 & 91.5 (mm) & 92.1 & 91.5 & 92.3 (mam) \\
RRM1 & constellation & 1 & 100.0 & 86.0 & 86.1 & 86.2 (pc) & 69.9 & 69.3 & 69.9 (ms) \\
pikachurin & constellation & 2 & 100.0 & 92.1 & 90.4 & 94.5 (ms) & 24.0 & 22.2 & 25.8 (ms) \\
Sonic hedgehog & constellation & 1 & 100.0 & 100.0 & 100.0 & 100.0 (mm) & 73.4 & 73.2 & 76.2 (mam) \\
p53 & constellation & 1 & 100.0 & 100.0 & 100.0 & 100.0 (mm) & 70.5 & 70.0 & 71.5 (ms) \\
\midrule
POLG2 mRNA & angle & 2 & 100.0 & 68.3 & 72.8 & 83.4 (ms) & 57.0 & 47.5 & 68.5 (ms) \\
SUR1 isoform 2 & angle & 2 & 94.5 & 16.4 & 26.9 & 31.1 (pc) & 11.3 & 11.4 & 17.4 (ms) \\
MT-CO3 & angle & 1 & 100.0 & 61.9 & 81.5 & 87.9 (ms) & 65.8 & 84.7 & 90.2 (ms) \\
RRM1 & angle & 1 & 100.0 & 24.7 & 46.2 & 47.6 (mam) & 24.2 & 39.9 & 40.5 (mam) \\
pikachurin & angle & 2 & 94.9 & 24.2 & 38.9 & 50.1 (pc) & 13.9 & 12.1 & 20.6 (pc) \\
Sonic hedgehog & angle & 1 & 100.0 & 39.8 & 64.1 & 66.0 (mam) & 31.6 & 46.1 & 46.5 (mam) \\
p53 & angle & 1 & 100.0 & 44.5 & 60.1 & 60.1 (mm) & 33.3 & 46.3 & 46.3 (mm) \\
    \bottomrule
  \end{tabular}
  \caption{Ratio of correctly decoded symbols to sequence length, at 32,768 shots per measurement. 
  Per slot: symbols stored in each slot. Best
  DRR: the dynamic range recovery method with the highest accuracy, chosen with the true
  sequence; it matches min and max (mm), two percentiles (pc), mean and spread (ms),
  the absolute maximum (am), or min, mean and max (mam).}
  \label{tab:genetic}
\end{table*}

All the DNA/RNA sequences and protein amino acid sequences were run both in simulation and on hardware at 32,768 shots per experiment. This is the minimum power of two at which the noiseless simulations return no statistical error on the assignment of symbols in the constellation encoding. The results are shown in Fig.~\ref{fig:genetic-results} and Table~\ref{tab:genetic}. The angle encoding greatly underperforms the constellation encoding for this class of data, especially when attempting to fill the capacity, where the recovery rate is only around $25$\%. When filling the capacity to half, the distance between points is much larger and even the densely packed protein sequences are recovered at around $70$\%. The DNA/RNA sequences are much less densely packed, and their sequences are recovered with $80$\% symbols correct. All of these results fit well with the intuition that denser packing leads to higher susceptibility to noise and lower recovery rates. The large discrepancy in the two slot encoding for protein sequences is likely due to the simplified noise model not capturing a critical component of the noise. Another explanation is that the chance of an off by one symbol error is very high, and looking at the values directly would yield a more similar number. However, in biology even a single wrong nucleotide or amino acid can have deadly consequences, so it's not a useful metric. 

\end{document}